\documentclass[twocolumn]{svjour3} 

\usepackage{color}
\usepackage{graphicx}
\usepackage{amsmath}
\usepackage{amsfonts}
\usepackage{listings}
\usepackage[ruled]{algorithm}
\usepackage[noend]{algorithmic}
\usepackage{url}
\usepackage[]{todonotes}
\usepackage{verbatim}
\usepackage{xcolor}
\usepackage{multirow}
\usepackage{tabularray}
\usepackage{pifont}
\usepackage{xspace} 
\usepackage{tcolorbox}
\usepackage{colortbl}
\usepackage{subcaption}
\usepackage{makecell}
\usepackage{array}
\usepackage[table]{xcolor}
\usepackage{bbding, graphicx}
\usepackage{mdframed}
\usepackage{booktabs}
\usepackage{microtype}
\usepackage{fontawesome5}
\usepackage{float}

\usepackage[colorlinks=true, allcolors=blue]{hyperref}
\hypersetup{
    colorlinks=true,
    allcolors=blue
}

\definecolor{darkgreen}{RGB}{0,170,0}
\definecolor{darkred}{RGB}{200,0,0}
\definecolor{mygrey}{RGB}{230,230,240}
\definecolor{g5_c2}{HTML}{7ABBDB}
\colorlet{highcolor}{g5_c2!30!white}

\newcommand{\nop}[1]{}

\newcommand{\misunderstand}[1]{} 

\newenvironment{new}{\color{black}}{}

\newcommand{\proposal}{\textsc{LroBench}\xspace}
\newcommand{\correct}{{\color{darkgreen}\ding{52}}}
\newcommand{\wrong}{{\color{darkred}\ding{55}}}
\newcommand{\pcorrect}{{\color{cyan}\ding{52}\rotatebox[origin=c]{-9.2}{\kern-0.7em\ding{55}}}}
\newcommand{\LLMALL}{{\setlength{\fboxsep}{1pt}\colorbox{cyan!30}{\texttt{LLM-ALL}}}}
\newcommand{\LLMONE}{{\setlength{\fboxsep}{1pt}\colorbox{red!20!yellow}{\texttt{LLM-ONE}}}}
\newcommand{\LLMSEMI}{{\setlength{\fboxsep}{1pt}\colorbox{lime!70}{\texttt{LLM-SEMI}}}}

\newcommand\rong[1]{\textcolor{purple}{(Rong) #1}}
\newcommand\su[1]{\textcolor{cyan}{(Su) #1}}

\newcommand{\ie}{\emph{i.e.}\xspace}
\newcommand{\eg}{\emph{e.g.}\xspace}
\newcommand{\wrt}{\emph{w.r.t.}\xspace}
\newcommand{\aka}{\emph{a.k.a.}\xspace}
\newcommand{\vs}{\emph{vs.}\xspace}

\newcommand{\stitle}[1]{\noindent\underline{\textbf{#1}}}

\newcounter{insight}
\renewcommand{\theinsight}{\arabic{insight}}
\newcommand{\insight}[2]{ 
    \refstepcounter{insight} 
    \begin{tcolorbox}[
        boxrule=1pt, 
        left=0pt, right=0pt, top=0pt, bottom=0pt
    ]
    \label{#1} 
    \hspace{0.5em} 
    \textit{\textbf{I\theinsight}}\textbf{: } #2 
    \end{tcolorbox}
}
\newcommand{\iref}[1]{\hyperref[#1]{\textbf{I\ref{#1}}}}

\newcommand{\halfcorrect}{\Checkmark\kern-1.2ex\raisebox{1ex}{\rotatebox[origin=c]{125}{\textbf{--}}}}

\newlength{\figwidths}
\newlength{\figwidthd}
\newlength{\expwidths}
\newlength{\expwidthd}
\newlength{\legendwidths}
\begin{document}

\title{Large Language Model-Enhanced Relational Operators: \break Taxonomy, Benchmark, and  Analysis}

\author{Yunxiang Su$^{1,2,*}$, Tianjing Zeng$^{1,3,*}$, Zhongjun Ding$^{1}$, Yin Lin$^{1}$, \\ 
Rong Zhu$^{1,\#}$, Zhewei Wei$^{3}$, Bolin Ding$^{1}$, Jingren Zhou$^{1}$
}
\institute{
Yunxiang Su \at
\email{suyunxiang.syx@alibaba-inc.com}     
\and
Tianjing Zeng \at
\email{zengtianjing.ztj@alibaba-inc.com}  
\and
\faEnvelope  Rong Zhu \at
\email{red.zr@alibaba-inc.com}
\and
$^{1}$Alibaba Group, Beijing, China,
$^{2}$Tsinghua University, Beijing, China,
$^{3}$Renmin University of China, Beijing, China, 
$^{*}$Equal contribution, 
$^{\#}$Corresponding author 
}
\date{Received: date / Accepted: date}

\titlerunning{Large Language Model-Enhanced Relational Operators: Taxonomy, Benchmark, and  Analysis}
\authorrunning{Su et al.}

\maketitle
\begin{abstract}
With the development of large language models (LLMs), 
numerous studies integrate LLMs through operator-like components to enhance relational data processing tasks, 
including 
semantic filtering, knowledge-augmented table imputation, 
reasoning-driven entity matching, and more challenging semantic query processing. 
These components invoke LLMs while preserving a relational input/output interface, 
which we refer to as \emph{LLM-Enhanced Relational Operators (LROs)}.
From an operator perspective, unfortunately, 
these existing LROs suffer from fragmented definitions, various implementation strategies and inadequate evaluation benchmarks. 
To this end, in this paper, we first establish a unified LRO taxonomy to align existing LROs, and categorize them into five basic types: \textsf{Filter}, \textsf{Match}, \textsf{Impute}, \textsf{Cluster} and \textsf{Order}, along with their operand granularities and implementation variants.
Second, we design \textsc{LroBench}, a comprehensive benchmark featuring 290 single-LRO queries and 60 multi-LRO queries, spanning 27 databases across more than 10 domains.
\textsc{LroBench} covers all operating intents and operand granularities in its single-LRO workload, and provides challenging multi-LRO queries stratified by query complexity. 
Based on these, we evaluate individual LROs under various implementations, 
deriving practical insights into LRO design choices and summarizing our empirical best practices.
We further compare the end-to-end performance of existing multi-LRO systems against an LRO suite instantiated with these best practices, 
in order to investigate how to design an effective LRO set for multi-LRO systems targeting complex queries.
Last, to facilitate future work, we outline promising future directions and open-source all benchmark data and evaluation code, available at \url{https://github.com/LROBench/LROBench/}.
\end{abstract}

\section{Introduction}

\nop{
The emergence and evolution of large language models (LLMs) have inspired efforts and advancements in improving database query processing, 
mainly through the semantic reasoning capabilities and world knowledge of LLMs.
}

The emergence and evolution of large language models (LLMs) have inspired numerous efforts and significant advancements in improving relational data processing.
For example, LLM-enhanced entity matching~\cite{comem,fm4dt,batcher} leverages LLMs to judge whether tuples across two tables refer to the same real-world entity, and 
prompt-based table imputation~\cite{unidm,DBLP:conf/vldb/ZhangD0O24,DBLP:conf/acl/HeBZWCH25} uses LLMs to reason about and fill missing cells.
Moreover, 
semantic query processing engines~\cite{sembench,blendsql,lotus,tag,palimpzest,abacus} (see formal definition in SemBench~\cite{sembench}) extend relational algebra and SQL with semantic operators that invoke LLMs to evaluate natural-language (NL) predicates during operator execution.
LLMs are essential for these tasks as they enable world-knowledge retrieval, semantic reasoning and generation.

Despite targeting different tasks, 
a substantial portion of these systems and methods integrate LLMs through operator-like components
(\eg, \texttt{sem\_filter} in LOTUS-TAG~\cite{lotus,tag}, \texttt{LLMJoin} in BlendSQL~\cite{blendsql}).
These components share a common pattern:
(i) operate over relational data, 
(ii) take an NL requirement (\ie, instruction or condition), 
(iii) invoke LLMs to make semantic predicates during execution, and 
(iv) produce relational outputs.
Motivated by this trend, we refer to these components as \emph{LLM-Enhanced Relational Operators (LROs)}, 
formally defined in Section~\ref{sec:pre-lro}.

Research on LROs is progressing rapidly. 
Recent studies have proposed a variety of LROs~\cite{DBLP:conf/edbt/Katsogiannis-Meimarakis26,c3,fm4dt,comem,unidm,DBLP:conf/vldb/ZhangD0O24,llm-gdo,DBLP:conf/icde/HuangW24,batcher,DBLP:conf/acl/HeBZWCH25}, 
as well as semantic query processing engines~\cite{sembench} that expose suites of LROs~\cite{binder,suql,aryn,hqdl,thalamusdb,blendsql,palimpzest,abacus,docetl,lotus,tag}.
Yet these LROs have not been systematically reviewed, logically aligned or comprehensively evaluated, 
leaving a fragmented landscape. We identify three gaps.
First, operators with different names (and even different operands) may share the same underlying intent, but such intents are rarely made explicit and aligned (see Q1, Section~\ref{sec:pre-roadmap}). For example, \texttt{NLJoin} in ThalamusDB~\cite{thalamusdb} and \texttt{Merge} in DocETL~\cite{docetl} both aim to semantically join tables. 
Second, even for the same LRO, systems adopt different implementation variants and design choices, yet these are seldom compared in a unified, standalone (single-LRO) evaluation (see Q2, Section~\ref{sec:pre-roadmap}). 
For example, nested-loop join in LOTUS-TAG~\cite{lotus,tag} versus semi-join in BlendSQL~\cite{blendsql}. 
Third, real applications rely on pipelines that integrate multiple LROs, where optimizing one operator in isolation may not improve end-to-end performance. 
It thus remains unclear whether existing multi-LRO systems make effective integrated design choices, and how to design an effective LRO suite for such systems (see Q3, Section~\ref{sec:pre-roadmap}).
\emph{To bridge these gaps, there is a pressing need for a comprehensive benchmark on LROs from both operator and system-level perspectives.}


Unfortunately, there is currently no benchmark specifically designed for LRO evaluation. 
Moreover, existing benchmarks, even those most relevant to the LRO scope, still fall short of our requirements.
Specifically, fact check~\cite{fever,tablefact,feverous} and question answering benchmarks~\cite{hybridqa,ott-qa,mmqa,tqa-bench} do not preserve a relational input/output interface. 
Text-to-SQL benchmarks~\cite{bird,spider,wikisql,beaver,spider2.0,elt-bench} do not consider or involve any LROs, 
but only use LLMs to translate NL questions into classical SQLs.
Other prior benchmarks on semantic query processing, like TAG~\cite{tag}, SWAN~\cite{hqdl} and SemBench~\cite{sembench}, 
evaluate the end-to-end performance of certain multi-LRO systems. 
However, none of them drill down to individual LROs, let alone their operating intents or implementation variants.
Such a lack of an operator-level view also impedes deeper system-level analysis
(see Section~\ref{sec:bench-rw}).

\begin{figure*}[t]
    \centering
    \includegraphics[width=0.95\figwidthd]{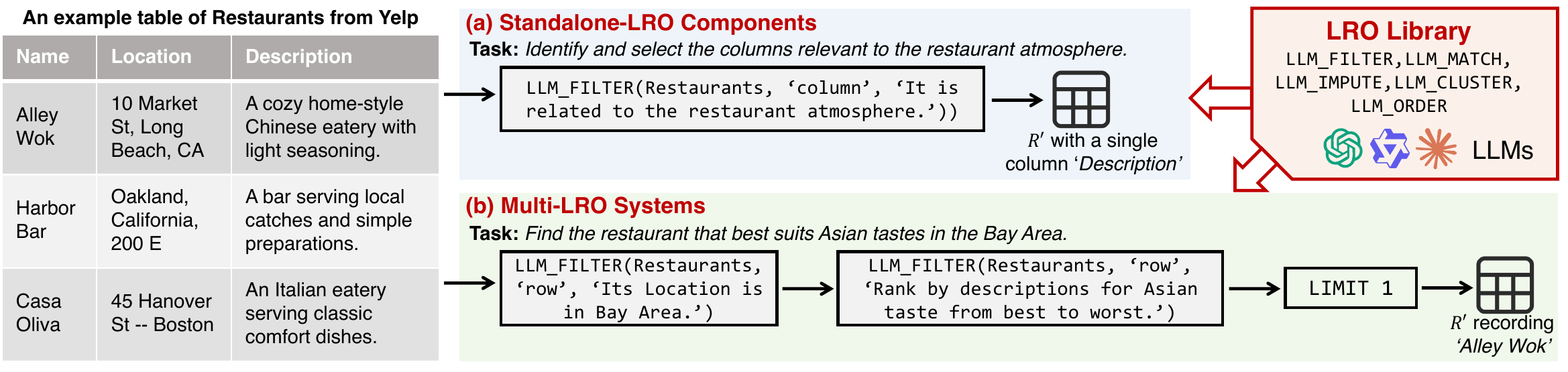}
    \caption{Examples of standalone-LRO components and multi-LRO systems.}
    \label{fig:system-example}
\end{figure*}

Therefore, in this paper, we aim to taxonomize, benchmark, evaluate and analyze LROs, progressing from the operator level to the system level. 
Our contributions are summarized as follows.

\emph{\textbf{(1) A unified LRO taxonomy in Section~\ref{sec:op-tax}. }}%
To align existing LROs,
we establish the \emph{first unified LRO taxonomy} in Table~\ref{tab:op-tax}.
We categorize LROs by the intent of LLM request, 
and group them into five basic types: \textsf{Filter}, \textsf{Match}, \textsf{Impute}, \textsf{Cluster} and \textsf{Order}.
Then, we further categorize them along 
operand granularity (\ie, cell, row, column, or table), 
and implementation variant (\eg, \texttt{LLM-ALL} with a batched LLM call, or \texttt{LLM-ONE} invoking the LLM iteratively).

\emph{\textbf{(2) A comprehensive benchmark for LROs in Section~\ref{sec:bench}. }}%
To evaluate LROs at both operator and system levels,  
we design a comprehensive LRO benchmark \proposal, 
spanning 27 real-world databases across more than 10 domains.
It includes 290 single-LRO queries covering all LRO types and operands in our taxonomy,  
filling the gap left by existing benchmarks.
It also includes 60 multi-LRO queries stratified by complexity, featuring high diversity.

\emph{\textbf{(3) Evaluation of individual LRO in Section~\ref{sec:op-eval}.}}
To assess implementation variants of each LRO, we conduct evaluation on individual LROs in terms of effectiveness and scalability.
We derive: (i) \emph{concrete LRO design choices for each LRO} in \iref{1}--\iref{5}, summarized as best practices in Table~\ref{tab:best-practices}, 
and (ii) \emph{LRO design trade-offs} in \iref{6}--\iref{7}.

\emph{\textbf{(4) Evaluation of multi-LRO systems and best practices in Section~\ref{sec:e2e-eval}.} }
To study how to build advanced multi-LRO systems,
we evaluate current multi-LRO systems against our best practices.
Remarkably, our best-practice baseline achieves the strongest overall performance, reaching up to 86.67\% correctness.
We analyze: (i) \emph{early-stage landscape and best-practice guidelines} in \iref{8}--\iref{9},
(ii) \emph{system performance factors: LRO support, LRO design, base model} in \iref{10}--\iref{11},
and (iii) \emph{failure modes} in \iref{12}.

\nop{
Key findings (marked as \iref{8} to \iref{15}) are highlighted as follows:

\underline{$\bullet$ {\bf K4:}} 
Foundational principles for enhancing effectiveness (\iref{8}) and scalability (\iref{13}, \iref{14}).

\underline{$\bullet$ {\bf K5:}}
Operator-specific optimization strategies: semantic matching accuracy (\iref{9}), imputation quality (\iref{10}), clustering consistency (\iref{11}), and ordering reliability (\iref{12}).

\underline{$\bullet$ {\bf K6:}} 
Context-aware adaptation principles for scalable operator execution (\iref{15}).
}

\emph{\textbf{(5) Conclusion and future directions in Section~\ref{sec:conclusion}.}}
We conclude our efforts and 
outline future directions based on our findings, 
including (i) \emph{well-designed LRO systems} with full coverage, robust planning, joint optimization and optimal implementations in the short term, and 
(ii) \emph{LLM-centric databases} in future.

\section{Background}
\label{sec:pre}

\subsection{LLM-Enhanced Relational Operator}
\label{sec:pre-lro}
A conventional relational algebra operator processes database relations: it accepts one or two relations as input and produces a relation as output~\cite{DBLP:books/daglib/0006733}. 
We extend this notion to an LLM-enhanced relational operator (LRO), 
defined as follows.

\begin{definition}[\underline{\textbf{L}}LM-Enhanced \underline{\textbf{R}}elational
\underline{\textbf{O}}perator, \textbf{LRO}]
\label{def:lro}
An LLM-enhanced relational operator (LRO) is a logical extension of SQL that enables operations beyond the capability of traditional SQL. 
Formally, 
an LRO takes as input one or two relations $\mathcal{R} \in \{\{R\}, \{R_1, R_2\}\}$, an operand granularity $g \in \{\textsf{cell},\textsf{row},\textsf{column},\textsf{table}\}$, and an NL requirement $l$, 
and produces an output relation $R'$:
\[
\mathsf{LRO}(\mathcal{R}, g, l) \mapsto R',
\]
where the LRO treats the input relation(s) as cells, rows, columns, or tables according to $g$, 
invokes an LLM to transform the input according to $l$, 
and materializes the result as a new relation $R'$.
\end{definition}%

Notably, LROs are closely related to \emph{semantic operators} defined in SemBench~\cite{sembench}, 
as both leverage LLMs to capture and generate semantic information beyond purely syntactic relational processing.
However, they slightly differ in scope and design: semantic operators typically target heterogeneous data domains (\eg, images or documents), whereas LROs focus strictly on relational data; 
LROs further emphasize LLM-based reasoning,  numerical computation and  external knowledge retrieval, 
and operate at multiple granularities (cells, rows, columns, or tables).

\begin{example}
Figure~\ref{fig:system-example} illustrates three LRO examples. In Figure~\ref{fig:system-example}(a), \texttt{LLM\_FILTER} takes as input one relation with restaurant records, 
performs semantic schema linking (\ie, selects the columns relevant to the restaurant atmosphere), 
and outputs a relation whose schema is narrowed to include only the ``Description'' column.
In Figure~\ref{fig:system-example}(b), \texttt{LLM\_FILTER} filters rows in the relation by leveraging the LLM’s external knowledge to determine whether a restaurant is located in the Bay Area. 
Then, \texttt{LLM\_ORDER} applies the LLM’s semantic reasoning to rank the restaurants by their taste. 

\begin{new}
Note that we name LROs by the operating intent of the LLM request, \ie, what the LLM is asked to do, or the action specified by the prompt during LRO execution. 
Although \texttt{LLM\_FILTER} in Figure~\ref{fig:system-example}(a) performs a column projection while \texttt{LLM\_FILTER} in Figure~\ref{fig:system-example}(b) filters rows, 
they both use the LLM to carry out a filtering intent and therefore fall into the same LRO type and name (see Section~\ref{sec:op-tax} for other intents).
\end{new}%
\end{example}

\subsection{Relevant Works on LROs}
\label{sec:pre-mt}
\begin{new}
Numerous existing works have applied LLMs to relational data. In this paper, we focus on studies whose LLM-enhanced components fall within the scope of LROs following Definition~\ref{def:lro}. 
While LROs are in principle composable with each other and with classical SQL operators, prior works differ in how they are packaged and exposed.
Accordingly, we categorize existing works into two groups.
\end{new}%

\textbf{Standalone-LRO works} expose a single LRO as an end-to-end component for a relatively self-contained task.
LROs in these works typically target data preparation (\eg, entity matching) over relations, and therefore often provide limited support for composing the LRO with SQL operators or additional LROs. 
For example, Figure~\ref{fig:system-example}(a) shows \texttt{LLM\_FILTER} at the column level, which performs end-to-end schema linking and returns a narrowed relation.

The related literature on standalone-LRO components is extensive, and we only review representative ones. 
The LLM-enhanced schema linking methods in \underline{LLM-SL}~\cite{DBLP:conf/edbt/Katsogiannis-Meimarakis26}, \underline{C3}~\cite{c3} and others~\cite{DBLP:conf/coling/LiuTZXZWHL25,DBLP:conf/aaai/Li00023,DBLP:conf/acl/GanCP23} can be cast as a standalone LRO that selects semantically relevant columns and tables.
In addition, current prompt-based entity matching methods~\cite{fm4dt,comem,unidm,batcher,DBLP:conf/edbt/PeetersSB25,DBLP:conf/www/LiLHZSC24} can also be cast as an LRO that matches tuples across two relations and outputs a relation of matched tuple pairs,
where entity matching components in \underline{DataWrangle}~\cite{fm4dt} and \underline{ComEM}~\cite{comem} are representative examples.
Table imputation components~\cite{fm4dt,unidm,DBLP:conf/acl/HeBZWCH25,DBLP:conf/vldb/ZhangD0O24} can also be cast as an LRO, since they take an incomplete table as input and produce a filled table.
Imputation components in \underline{UniDM}~\cite{unidm} and \underline{LLM-Preprocessor}~\cite{DBLP:conf/vldb/ZhangD0O24} are considered as representative ones.
Last, new column inference and generation methods~\cite{DBLP:conf/icde/HuangW24,fm4dt,unidm,DBLP:journals/pacmmod/LiHYCGZF0C24,llm-gdo} are also LROs that add a new column to a relation.
Exemplars include \underline{LLM-GDO}~\cite{llm-gdo} and \underline{LLM-Relationalization}~\cite{DBLP:conf/icde/HuangW24}.

\textbf{Multi-LRO systems} provide a composable suite of LROs that can be nested within each other and integrated with classical SQL operators to solve complex tasks, 
\begin{new}
typically for semantic queries, 
which refer to database queries based on the meaning and semantics of the data rather than exact value matching alone.
\end{new}%
For example, Figure~\ref{fig:system-example}(b) illustrates a multi-LRO system that solves a semantic query beyond traditional SQL capabilities.
The solution composes two LROs, \texttt{LLM\_FILTER} and \texttt{LLM\_ORDER} at the row level, with the classical \texttt{LIMIT} operator in a single query plan.

Notably, these systems usually orchestrate LROs and SQL operators into a query plan, obtained either via automated LLM-based planning or through manual annotation.
\underline{Binder}~\cite{binder} augments SQL with two operators: \texttt{f\_col} for column generation and \texttt{f\_val} for non-relational summarization (not an LRO). 
\underline{SUQL}~\cite{suql}, designed for conversational agents and validated on Yelp datasets, 
provides \texttt{Answer} (cell/column responses) and \texttt{Summary} (free-text generation, not an LRO). 
Besides, \underline{Aryn-Sycamore}~\cite{aryn} can automatically build a plan to analyze a document set,  
while being also compatible with relations by treating each tuple as a document.
It supports two LROs on tables: \texttt{Map} for new column generation by inferring from the table, 
and \texttt{Filter} for semantic filtering. 
Other LLM operators in Aryn-Sycamore are not compatible with relations, thus excluded.

Other systems rely on manually annotated plans.
\underline{HQDL}~\cite{hqdl} proposes \texttt{Generate} for semantic column generation,
which can be seamlessly integrated with SQL.
\underline{ThalamusDB}~\cite{thalamusdb} designs \texttt{NLFILTER} and \texttt{NLJOIN} for semantic selection and joining, respectively. 
\underline{BlendSQL}~\cite{blendsql} extends SQL with three LROs: \texttt{LLMValidate} (filtering), 
\texttt{LLMMap} (similar to HQDL's \texttt{Generate}), and \texttt{LLMJoin}. 
\underline{Palimpzest}~\cite{palimpzest,abacus} proposes core LROs: 
\texttt{Map}, \texttt{Filter}, and \texttt{Join}. 
Although \underline{DocETL}~\cite{docetl} primarily targets document analytics, its LROs 
\texttt{Map}, \texttt{Filter}, and \texttt{Merge} (semantic joining) resemble Palimpzest's operators and are compatible with relations.
\underline{LOTUS-TAG}~\cite{lotus,tag} offers the most comprehensive suite with five LROs: 
\texttt{sem\_filter}, \texttt{sem\_join}, \texttt{sem\_topk}, \texttt{sem\_map}, and \texttt{sem\_cluster\_by}.

\stitle{Exclusion.}
We exclude several lines of work that fall outside the LRO scope in Definition~\ref{def:lro}.
First, methods that only process unstructured data rather than relational data are excluded, 
such as ZenDB~\cite{zendb}, QUEST~\cite{quest} and CAESURA~\cite{caesura}. 
Second, Text-to-SQL methods  are excluded, since they only use classical SQLs without LLM-enhanced relational operators, 
such as Din-SQL~\cite{din-sql}, MAC-SQL~\cite{mac-sql} and CHASE-SQL~\cite{chase-sql}.
Third, techniques that rely  on embeddings or fine-tuning are excluded, 
as they mainly perform similarity search over vector data rather than invoking an LLM for a semantics-aware transformation over relations.
Examples include vanilla RAG-based methods~\cite{DBLP:conf/nips/LewisPPPKGKLYR020}, the \texttt{index}, \texttt{search} operators in LOTUS-TAG, and the \texttt{Top-K} operator in Palimpzest (which retrieves data from vector databases by embedding similarity).
Last, LLM operators for free-text generation or table summary are excluded,
since they do not materialize outputs as relations.
Examples include \texttt{f\_{val}} in Binder, \texttt{Summary} in SUQL, \texttt{LLMQA} in BlendSQL, \texttt{Aggregate} in Palimpzest, \texttt{Agg} in DocETL, and \texttt{sem\_agg} in LOTUS-TAG.

\subsection{Motivation and Roadmap}
\label{sec:pre-roadmap}

Existing LRO-related systems exhibit  diverse operator patterns.
However, despite using different names, 
certain operators across different systems share the same underlying operating intent. 
For instance, both \texttt{NLJoin} in ThalamusDB~\cite{thalamusdb} and \texttt{Merge} in DocETL~\cite{docetl} join relations by semantic comparison of join keys.
Moreover, operators targeting different operands may also follow the same intent.
For example, ComEM~\cite{comem} can be cast as a standalone LRO for row-level entity matching, logically similar to cell-level join operators such as \texttt{NLJOIN} and \texttt{Merge}.
This raises our first question: \textbf{\emph{How many distinct operating intents underlie current LROs?}} (\textbf{Q1})

Furthermore, LROs with the same name, intent and operand may be implemented under various strategies across systems.
For instance, both LOTUS-TAG~\cite{lotus,tag} and BlendSQL~\cite{blendsql} support semantic join operations.
However, LOTUS-TAG follows a nested-loop strategy that prompts one pair of values per LLM invocation, 
whereas BlendSQL adopts a left-join strategy that prompts one left value together with all right values, \aka, a semi-join.
These various implementations induce different prompt contexts and reasoning patterns, 
which can affect the accuracy of LLM judgments.
Thus, we raise the second question: \textbf{\emph{How do different LRO implementations impact their performance?}} (\textbf{Q2})

Finally, while Q1--Q2 focus on the intents and implementations of individual LROs,
real-world applications, especially semantic query processing engines~\cite{sembench}, are supported by multi-LRO systems with an LRO set.
In such systems, designing a multi-LRO solution requires going beyond optimizing a single LRO in isolation, 
and instead understanding how to incorporate multiple LROs into a coherent end-to-end system, each implemented using operator-level best practices.
This raises our third question: \textbf{\emph{How should multi-LRO systems be designed to incorporate operator-level best practices, and do existing systems do so effectively?}} (\textbf{Q3})

\stitle{Roadmap.}
For \textbf{Q1}, we develop a unified taxonomy of LRO operating intents to systematically categorize and align diverse LLM-enhanced operators (Section~\ref{sec:op-tax}).
For \textbf{Q2}, we build a comprehensive benchmark with single-LRO queries to evaluate individual LROs (Section~\ref{sec:bench}), 
and compare various implementations of each LRO type (Section~\ref{sec:op-eval}).
For \textbf{Q3}, we extend the benchmark with multi-LRO queries (Section~\ref{sec:bench}) and evaluate current multi-LRO systems against our best practices derived from our taxonomy and operator-level insights (Section~\ref{sec:e2e-eval}).
Together, these results show that our taxonomy and implementation insights inform multi-LRO system design and yield measurable end-to-end gains.

\section{Operator Taxonomy}
\label{sec:op-tax}

\begin{table*}[t]
    \centering
    \small
    \caption{A taxonomy of LLM-enhanced relational operators.}
    \label{tab:op-tax}
    \resizebox{\linewidth}{!}{%
    \begin{tblr}{
      cell{2}{1} = {r=3}{},
      cell{5}{1} = {r=3}{},
      cell{8}{1} = {r=3}{},
      cell{11}{1} = {r=3}{},
      hline{1,15} = {-}{0.12em},
      hline{2,5,8,11,14} = {-}{0.05em},
      hline{3-4,6-7,9-10,12-13} = {2-6}{}, 
      rows = {valign=m},
      row{1} = {bg=mygrey, font=\bfseries}
    }
    \textbf{Intent} & \textbf{Operand} & \textbf{Functionality} & \textbf{Primitive} & \textbf{Existing LROs and Implementations} \\
    \textsf{Filter}
        & Row  & {Filter records by semantic conditions} & {Select ($\sigma$)} & {
        \LLMALL $\:$ BlendSQL: \texttt{LLMValidate} \\
        \LLMONE $\:$ Aryn-Sycamore: \texttt{Filter}  $\: \mid$ ThalamusDB: \texttt{NLFilter} \\
        Palimpzest: \texttt{Filter} $\: \mid$
        DocETL: \texttt{Filter} $\: \mid$ LOTUS-TAG: \texttt{sem\_filter} } \\
        & Column & {Find columns by semantic conditions} & {Project ($\pi$)} & { \LLMALL $\:$  LLM-SL$\: \mid$ C3 $\:$ 
        \LLMONE $\:$ ---} \\
        & Table & {Identify tables by semantic conditions} & {---}
        & {\LLMALL $\:$ LLM-SL$\: \mid$ C3 $\:$ 
        \LLMONE $\:$ --- } \\
    \textsf{Match}
        & Cell & {Establish a semantic matching  across \\ two sets of cells to support table joining} & {Join ($\bowtie$)} & {
        \LLMALL $\:$
         ThalamusDB: \texttt{NLJoin} (Batch)   \\
        \LLMONE $\:$  
        ThalamusDB: \texttt{NLJoin} (Nested-Loop)  $\: \mid$ 
        Palimpzest:\texttt{Join} \\ DocETL: \texttt{Merge}  \\
        \LLMSEMI $\:$ BlendSQL: \texttt{LLMJoin}  $\: \mid$ 
        LOTUS-TAG: \texttt{sem\_join} } \\
        & Row & {Find records with semantic connections, \\ such as entity resolution (matching) } & {---} & {
        \LLMALL $\:$ ComEM ($m$:$n$) $\:$ 
        \LLMONE $\:$ DataWrangle $\: \mid$ ComEM ($1$:$1$) \\
        \LLMSEMI $\:$ ComEM ($1$:$n$)} \\
        & Column & {Identify semantically joinable columns \\ on different tables} & {---} &
        \LLMALL $\:$ --- $\:$ \LLMONE $\:$ --- $\:$ \LLMSEMI $\:$ ---  \\
    \textsf{Impute}
    & Cell & {Infill missing cell values} & {---} & {
        \LLMALL $\:$ --- $\:$ 
        \LLMONE $\:$  
        LLM-Preprocessor $\: \mid$ 
        UniDM:\texttt{Imputation} } \\
    & Row &  {Insert new records into databases} & {---} & {
         \LLMALL $\:$ --- $\:$ \LLMONE $\:$ ---
        } \\
    & Column & {Extend tables with new columns with \\ additional information} & {---} & {
        \LLMALL $\:$  Binder: \texttt{f\_{col}} $\: \mid$ 
        BlendSQL: \texttt{LLMMap} $\: \mid$ 
        UniDM: \texttt{Transform} \\  LLM-Relationalization
         \\
        \LLMONE $\:$ SUQL: \texttt{Answer}  $\: \mid$ 
        Aryn-Sycamore: \texttt{Map}  $\: \mid$ 
        HQDL: \texttt{Generate} \\
        Palimpzest: \texttt{Map}  $\: \mid$
        DocETL: \texttt{Map} $\: \mid$
        LOTUS-TAG: \texttt{sem\_map} $\: \mid$
        LLM-GDO
    } \\
    \textsf{Cluster}
    & Row  & {Cluster records by semantic conditions} & {GroupBy ($\gamma$)} & { \LLMALL $\:$ LOTUS-TAG: \texttt{sem\_cluster\_by} $\:$
    \LLMONE $\:$ ---} \\
    & Column & {Organize columns by semantics, \\ enabling  table segmentation} & {---} & \LLMALL $\:$ --- $\:$  \LLMONE $\:$ ---    \\
    & Table & {Organize tables by semantics, serving \\ schema hierarchies} & {---} &
    \LLMALL $\:$ --- $\:$ \LLMONE $\:$ ---       \\
    \textsf{Order} & Row & {Rank records by knowledge specified \\ in semantic metrics} & {OrderBy ($\tau$)} &  {
    \LLMALL $\:$ --- $\:$
    {\setlength{\fboxsep}{1pt}\colorbox{red!20!yellow}{\texttt{LLM-PAIR}}} LOTUS-TAG: \texttt{sem\_topk} (naive) \\
    {\setlength{\fboxsep}{1pt}\colorbox{brown!50}{\texttt{LLM-SORT}}} $\:$ LOTUS-TAG: \texttt{sem\_topk} (heap) $\:$
    {\setlength{\fboxsep}{1pt}\colorbox{magenta!25}{\texttt{LLM-SCORE}}} $\:$ ---
    }\\
    \end{tblr}
    }
\end{table*}

\nop{
To resolve this issue, we consider how to reasonably align and categorize such operators in this section. We begin with a systematic operator taxonomy (in Section~\ref{sec:op-tax-overview}), where each type of operator is formally defined (in Section~\ref{sec:op-def}) and logically independent of each other. Their collection could cover the full database query spectrum. 
Based on this, we exhibit how existing LLM-enhanced database processing methods align with our standard operator taxonomy (in Table~\ref{tab:op-tax} and explained Section~\ref{sec:op-def}). This analysis provides a unified basis to compare these methods and lays a fair foundation for the further benchmarking and evaluation.}

To investigate Q1, we first generalize existing LROs in Section~\ref{sec:pre-mt} and establish a unified  taxonomy to categorize and align them.
Our taxonomy is organized along three dimensions to ensure its completeness and mutual exclusivity: 
\begin{new}
(1) \emph{Operating Intent}, \ie, what the LLM is asked to
do, or the action specified by the prompt;
\end{new}%
(2) \emph{Operand Granularity}, \ie, the level at which the LLM operates: cells, rows, columns, or tables, as defined in Definition~\ref{def:lro};
and (3) \emph{Implementation Variant}, \ie, the strategies used to realize an LRO, particularly how data are structured and fed into the LLM.
Table~\ref{tab:op-tax} illustrates our proposed taxonomy.

First, we categorize LROs by their operating intents into five mutually exclusive types (shown in the first column of Table~\ref{tab:op-tax}): \textsf{Filter}, \textsf{Match}, \textsf{Impute}, \textsf{Cluster} and \textsf{Order}, hereafter referred to as basic LRO types, each specifying the task or action assigned to the LLM.
Then, each LRO type is further classified by operand granularity (second column), 
where we list common combinations and omit rare ones (\eg, \textsf{Filter} + Cell, \textsf{Match} + Table).
We provide a brief description of each LRO (third column), and illustrate the relational algebra primitive of each LRO if possible (fourth column).
We then align all the LROs reviewed in Section~\ref{sec:pre-mt} to our taxonomy (fifth column).
Implementations are illustrated using colored rectangles, \eg, \LLMALL, \LLMONE. 
The symbol ``---'' implies that an implementation method is applicable to an operator yet remains unexplored in existing works.
We detail the operating intent, operand granularity and various implementations of the five LRO types below.

\stitle{LRO 1: \textsf{Filter}.}
\label{sec:op-select}
The enhanced \textsf{Filter} extends traditional SQL's \texttt{SELECT} (row filtering) and \texttt{PROJECT} (column selection) by retrieving subsets satisfying semantic conditions. Formally, we have

\begin{definition}[\textnormal{\textsf{Filter}}]
Given a relation $R$, an operand granularity $g\in\{\textsf{row}, \textsf{column}, \textsf{table}\}$ and an NL requirement $l$,
\textnormal{\textsf{Filter}} returns a subset of rows, columns or tables in $R$ (according to $g$) that satisfy $l$, \ie,
\[
\textnormal{\textsf{Filter}}(R,g,l) =\{\, e \in E_g(R) \mid e \models l \,\},
\]
where $E_g(R)$ denotes the set of rows, columns or tables obtained from $R$ at granularity $g$; the notation is used throughout.
\end{definition}%

\begin{figure}[t]
    \centering
    \includegraphics[width=\figwidths]{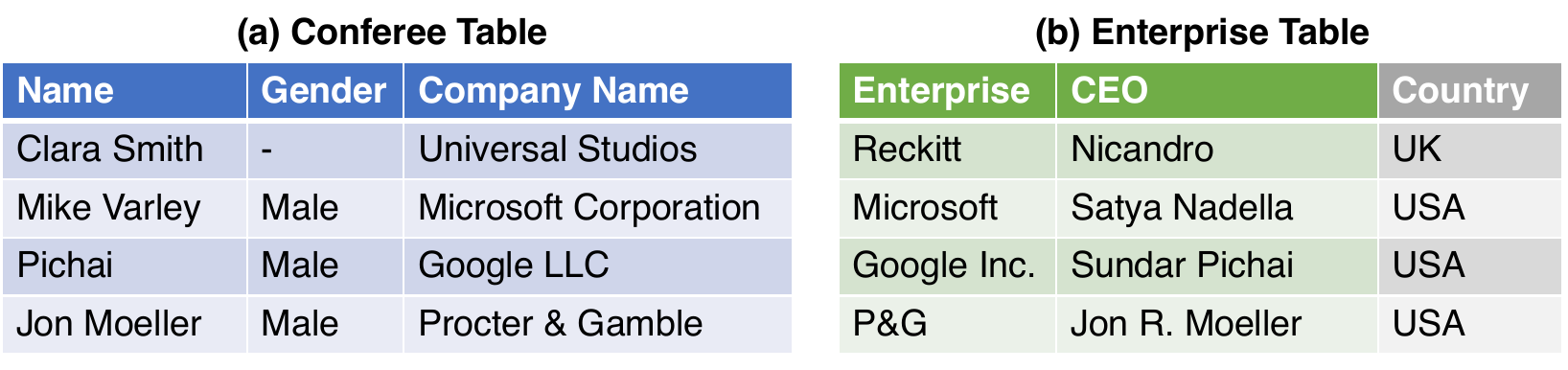}
    \caption{An example to illustrate LLM-enhanced operators.}
    \label{fig:example}
\end{figure}

The functionality of \textsf{Filter} varies by operand granularity. Row-wise \textsf{Filter} (\eg, \texttt{sem\_filter} in LOTUS-TAG~\cite{lotus,tag}) filters records. 
For example,  as shown in Figure~\ref{fig:example}, 
selecting ``technical companies'' from Table (b) returns Microsoft and Google. Column-wise and table-wise \textsf{Filter} identify relevant columns or tables, respectively. 
For instance, table-wise \textsf{Filter} for ``company CEO'' retrieves Table (b), and column-wise \textsf{Filter} further locates the ``CEO'' column.

There are mainly two implementations of \textsf{Filter}:
(1) \texttt{LLM-ALL} uses a batched strategy, which packs all candidates in $E_g(R)$ into a single prompt, submits it to the LLM, and invokes the LLM to directly return the filtered ones.
Currently, only BlendSQL~\cite{blendsql} applies this strategy.
(2) \texttt{LLM-ONE}, widely applied in current works~\cite{aryn,thalamusdb,palimpzest,abacus,docetl,lotus}, 
uses a candidate-wise strategy 
that individually packs each candidate in $E_g(R)$, queries the LLM with the semantic condition $l$, and returns a Boolean value (true/false).
After traversing all candidates, those labeled true are retained. 
\begin{new}
Given $n$ candidates, \texttt{LLM-ALL} and \texttt{LLM-ONE} respectively take $\mathcal{O}(1)$ and $\mathcal{O}(n)$ times of LLM calls.
\end{new}%

\stitle{LRO 2: \textsf{Match}.}
The enhanced \textsf{Match} operator generalizes SQL's \texttt{JOIN} operation beyond value-based matching to establish a semantic matching relationship between two sets of cells, rows or columns. Formally, we have,
\begin{definition}[\textnormal{\textsf{Match}}]
Given two relations $R_1$ and $R_2$, an operand granularity $g\in\{\textsf{cell},\textsf{row},\textsf{column}\}$  and an NL requirement $l$,
\textnormal{\textsf{Match}} returns a set of matched pairs of cells, rows or columns at granularity $g$ that satisfy $l$, \ie,
\[
\begin{aligned}
\textnormal{\textsf{Match}}&(R_1,R_2,g,l) = \\
&\{\, (e_1,e_2) \in E_g(R_1)\times E_g(R_2) \mid (e_1,e_2)\models l \,\}.
\end{aligned}
\]
\end{definition}%

Similarly, the functionality of \textsf{Match} also varies by operand types. Cell-wise matching enables semantic table joining (\eg, \texttt{LLMJoin} in BlendSQL~\cite{blendsql}). 
In Figure~\ref{fig:example}, Tables (a) and (b) can be joined on the condition ``the Company Name in Table (a) matches the Enterprise in Table (b)'' despite lacking primary-foreign keys. 
Row-wise matching identifies the same real-world entities, \eg, entity resolution in DataWrangle~\cite{fm4dt}.
Column-wise matching detects semantically joinable columns across tables (\aka schema matching in DataWrangle), exemplified by linking ``Company Name'' column in Table (a) and ``Enterprise'' column in Table (b).

We summarize three implementation strategies of \textsf{Match}: 
(1) \texttt{LLM-ALL} performs a one-time many-to-many  matching~\cite{thalamusdb,comem}, 
by submitting all candidates from both sets to the LLM in a single prompt and having it return all matches at once;
(2) \texttt{LLM-ONE} performs iterative one-to-one matching by enumerating candidate pairs and prompting the LLM to judge each pair individually~\cite{thalamusdb,palimpzest,abacus,docetl,lotus,fm4dt,comem}, 
\ie, an LLM-enhanced nested-loop join;
(3) \texttt{LLM-SEMI} performs one-to-many matching, that is, 
for each item in the left set, it prompts the LLM to select all matching items from the right set~\cite{blendsql,comem}, 
\ie, an LLM-enhanced semi-join or left join.
\begin{new}
Given $n$ candidates in a left table and $m$ candidates in a right table, \texttt{LLM-ALL}, \texttt{LLM-ONE} and \texttt{LLM-SEMI} take $\mathcal{O}(1)$, $\mathcal{O}(n*m)$ and $\mathcal{O}(n)$ times of LLM calls, respectively.
\end{new}%

\stitle{LRO 3: \textsf{Impute}.}
Inspired by SQL's \texttt{INSERT}, the LRO  \textsf{Impute} augments databases with LLM-generated content. 
We have,\begin{definition}[\textnormal{\textsf{Impute}}]
Given a relation $R$, an operand granularity $g\in\{\textsf{cell},\textsf{row},\textsf{column}\}$, and an NL  requirement $l$,
\textnormal{\textsf{Impute}} fills in the missing cells/rows/columns (according to $g$) of $R$ according to $l$ and returns an augmented relation $R'$, \ie,
\[
\textnormal{\textsf{Impute}}(R,g,l) = R',
\]
where imputed cells/rows/columns replace the corresponding missing ones in $E_g(R)$ and are materialized back to $R'$.
\end{definition}%

Cell-wise imputation populates missing values using LLM inference,
\eg, inferring ``Female'' for ``Clara'' in Table (a) of Figure~\ref{fig:example}, 
commonly applied in preprocessing tasks such as~\cite{unidm,DBLP:conf/vldb/ZhangD0O24}. 
Column-wise imputation appends new columns, exemplified by
\texttt{LLMMap} in BlendSQL~\cite{blendsql} and 
\texttt{Answer} in SUQL~\cite{suql}, 
for example, appending a new ``Country'' column to Table (b). 
Row-wise imputation inserts records using LLM knowledge or external sources, 
\eg, inserting a new row for ``OpenAI'' into Table (b).

Similar to \textsf{Filter}, there are two implementations of \textsf{Impute}: 
(1) \texttt{LLM-ALL}, which processes the entire table $R$ in one prompt~\cite{unidm,binder,blendsql,DBLP:conf/icde/HuangW24}, and 
(2) \texttt{LLM-ONE}, which imputes each element  individually~\cite{suql,aryn,hqdl,palimpzest,abacus,docetl,lotus,llm-gdo}.
\begin{new}
Similarly, 
with $n$ cells to impute, \texttt{LLM-ALL} needs $\mathcal{O}(1)$ LLM calls, whereas \texttt{LLM-ONE} needs $\mathcal{O}(n)$ LLM calls.
\end{new}%

\stitle{LRO 4: \textsf{Cluster}.}
Extended from SQL's \texttt{GROUPBY}, the 
LLM-enhanced \textsf{Cluster} considers clustering rows, columns or tables by semantics. That is,
\begin{definition}[\textnormal{\textsf{Cluster}}]
Given a relation $R$, an operand granularity $g\in\{\textsf{row},\textsf{column},\textsf{table}\}$, and an NL  requirement $l$,
\textnormal{\textsf{Cluster}} partitions the rows/columns/tables $E_g(R)$ into $k$ clusters guided by $l$:
\[
\textnormal{\textsf{Cluster}}(R,g,l) = \{C_1,\dots,C_k\},
\]
where the rows/columns/tables in $C_i$ belong to the same semantic cluster specified by $l$.
Besides, 
$\forall\, 1\le i\ne j\le k$, $C_i\cap C_j=\emptyset$, and $\bigcup_{i=1}^{k} C_i = E_g(R)$.
\end{definition}

Row-wise clustering enables semantic grouping of records, such as the \texttt{sem\_cluster\_by} operator in LOTUS-TAG. 
For example, it can cluster the enterprises in Table (b) into technology (Microsoft, Google) and retail (Reckitt, P\&G).
Column-wise clustering segments wide tables into sub-tables, \eg, 
grouping columns in Table (a) into basic information (Name, Gender) and job information (Company Name). 
Table-wise clustering organizes tables into schema hierarchies, such as grouping all personnel-related tables.

We consider two implementations: (1) \texttt{LLM-ALL}, which performs clustering in a single prompt with full visibility of all elements in $E_g(R)$~\cite{lotus}; 
and (2) \texttt{LLM-ONE}, which iteratively assigns each element in $E_g(R)$ to an existing cluster or creates a new one.  
\begin{new}
Given $n$ elements to cluster, \texttt{LLM-ALL} needs $\mathcal{O}(1)$ LLM calls, whereas \texttt{LLM-ONE} needs $\mathcal{O}(n)$ LLM calls.
\end{new}%

\stitle{LRO 5: \textsf{Order}.}
\label{sec:op-order}
The LLM-enhanced \textsf{Order} operator corresponds to the SQL's \texttt{ORDERBY}.
However, beyond ordering by numerical values, 
the enhanced 
\textsf{Order} further supports ordering by external knowledge and semantic criteria,
as defined below.
\begin{definition}[\textnormal{\textsf{Order}}]
Given a relation $R$ and an NL requirement $l$,
\textnormal{\textsf{Order}} (defined only at row granularity) sorts the rows in $E_{\textsf{row}}(R)$ according to $l$ and materializes the result as an output relation $R'$:
\[
\textnormal{\textsf{Order}}(R,\textsf{row},l) =  R'.
\]
\end{definition}

The enhanced \textsf{Order} operator exclusively operates at the row level, 
performing record ranking through qualitative commonsense knowledge 
that is hard to quantify, \eg, popularity and influence. 
The \texttt{sem\_topk} operator from LOTUS-TAG is a variation of the \textsf{Order} operator.
For example, in Figure~\ref{fig:example}, 
the \textsf{Order} can sort all the conferees in Table (a) by their social influence.

We summarize four implementations for \textsf{Order}: 
(1) \texttt{LLM-ALL}, which feeds all records into LLM in a single prompt and asks the LLM to complete the ordering directly;
(2) \texttt{LLM-PAIR}, which applies the LLM to compare each pair of records and then ranks each record by how many records surpass it~\cite{lotus,tag};
(3) \texttt{LLM-SORT}, which utilizes a quick sort but replaces the numerical comparison with LLM semantic comparison~\cite{lotus,tag};
and (4) \texttt{LLM-SCORE}, which first assigns a numerical score in [0, 100] to each record and then ranks them by the scores. 
Notably, for $n$ records, the four methods \texttt{LLM-ALL}, \texttt{LLM-PAIR}, \texttt{LLM-SORT} and \texttt{LLM-SCORE} invoke $\mathcal{O}(1)$, $\mathcal{O}(n^2)$, $\mathcal{O}(n\log n)$ and $\mathcal{O}(n)$ times of LLM calls, respectively. 
However, we find that implementations with a higher number of LLM calls do not necessarily perform better (see Section~\ref{sec:op-eval-order}). 

\begin{new}
\emph{Discussion on Operand Granularities.}
As in Definition~\ref{def:lro}, LROs may operate at four operand granularities: cell, row, column and table. 
In practice, however, providing an entire column or table as part of an LLM prompt is infeasible due to context-length limits. 
Therefore, for column and table granularities, we approximate the operand by including only the column/table name along with the first three cell/row values as a representative preview.
For cell and row granularities, the operands are typically small enough to fit within the prompt, so we provide the full information directly.
\end{new}

\section{Benchmark Design}
\label{sec:bench}

Given the LRO taxonomy in Section~\ref{sec:op-tax}, 
we design a novel benchmark \proposal to evaluate these LROs spanning different intents, operands and implementations. 
Section~\ref{sec:bench-goal} presents our design principles, Section~\ref{sec:bench-rw} surveys existing benchmarks, and Section~\ref{sec:bench-new} introduces \proposal with details.

\subsection{Design Principles}
\label{sec:bench-goal}

Aligning with our scope Q1-Q3 in Section~\ref{sec:pre-roadmap},
and the goal of benchmarking the LLM-enhanced relational operators defined in Definition~\ref{def:lro}, 
we establish the following design principles:

(1) \textbf{Native to Relations}: 
All benchmark queries should center on relational data, 
involve LLM-enhanced relational operators (\ie, \textsf{Filter}, \textsf{Match}, \textsf{Impute}, \textsf{Cluster} and \textsf{Order}), 
and operate on relational operands (\ie, cells, rows, columns and tables).

\begin{new}
(2) \textbf{Involvement of LROs}: 
The benchmark should ensure that each query explicitly involves one or more LROs (\ie, \textsf{Filter}, \textsf{Match}, \textsf{Impute}, \textsf{Cluster}, and \textsf{Order}) over relational operands (\ie, cells, rows, columns, and tables), 
rather than being solvable by purely classical SQL statements.

(3) \textbf{Complete Single-LRO Coverage in Isolation} (to address Q2):
The benchmark should include a dedicated set of single-LRO queries, each involving only one LRO, while collectively covering every LRO type and operand listed in Table~\ref{tab:op-tax}. 

(4) \textbf{Challenging Multi-LRO Composition} (to resolve Q3):
The benchmark should include more challenging multi-LRO queries that require composing multiple LROs within a task, 
enabling end-to-end evaluation of multi-LRO systems.
Moreover, these multi-LRO queries should be stratified into difficulty levels (\eg, LRO numbers) to reveal the capability limits of systems.
\end{new}

\vspace{-0.5em}
\subsection{Existing Benchmarks}
\label{sec:bench-rw}
\begin{new}
Unfortunately, there is currently no benchmark specifically designed for LRO evaluation. 
Moreover, existing benchmarks, even those most relevant to LRO scope, still fall short of our requirements.
\end{new}%
As in Table~\ref{tab:rw-bench}, they exhibit limitations on one or more dimensions, 
urging a more comprehensive benchmark on LROs.

(1) \underline{Fact check} benchmarks, \eg, FEVER~\cite{fever}, TableFact~\cite{tablefact} and FEVEROUS~\cite{feverous}, assess textual claims against evidence in structured tables, 
and return a yes or no verdict. 
Neither their inputs nor outputs are native to relations, 
let alone involving LROs.

(2) \underline{Question answering} benchmarks aim to answer declarative questions over relational tables.
HybridQA~\cite{hybridqa} and OTT-QA~\cite{ott-qa} design queries over single Wikipedia tables.
TQA-bench~\cite{tqa-bench} and MMQA~\cite{mmqa} propose more complex multi-hop questions over multiple tables. 
However, their outputs are unstructured responses rather than relations, 
and they do not involve LROs either.

(3) \underline{Text-to-SQL} benchmarks focus on translating declarative queries into formatted SQL statements. 
WikiSQL~\cite{wikisql} only considers simple SQL queries on single tables. 
Spider~\cite{spider} and BIRD~\cite{bird} expand to more domains, larger data volume and more complex SQL queries on multiple tables. 
BEAVER~\cite{beaver} collects  enterprise databases to improve real-world validity.
Spider 2.0~\cite{spider2.0} and ELT-Bench~\cite{elt-bench} further explore the translation to a SQL workflow. 
Their main efforts focus on aligning declarative queries with classical SQL operators through LLMs, 
but do not involve any LROs.

\begin{table}[t]
    \centering
    \caption{A review of existing benchmarks.}
    \label{tab:rw-bench}
    \resizebox{\figwidths}{!}{
    \begin{tabular}{r|cccc} 
    \toprule
     \rowcolor{mygrey}
     \textbf{Benchmark} 
     & \begin{tabular}[c]{@{}c@{}}\textbf{Native to}\\\textbf{Relations}\end{tabular} 
     & \begin{tabular}[c]{@{}c@{}}\textbf{Involvement}\\\textbf{of LROs}\end{tabular} 
     & \begin{tabular}[c]{@{}c@{}}\textbf{Single-LRO}\\\textbf{Isolation}\end{tabular}  
     & \begin{tabular}[c]{@{}c@{}}\textbf{Multi-LRO}\\\textbf{Composition}\end{tabular} \\ 
    \midrule
    Fact Check & \wrong & \wrong  & \wrong & \wrong \\
    QA Bench & \wrong & \wrong   & \wrong  & \wrong \\
    Text-to-SQL& \correct &  \wrong  & \wrong & \wrong \\
    SWAN~\cite{hqdl} & \correct & \correct & \wrong & \wrong \\
    TAG~\cite{tag} & \correct & \correct  & \wrong & \wrong \\ 
    SemBench~\cite{sembench} & \wrong & \correct & \pcorrect & \pcorrect \\
    \cellcolor{highcolor}
    \textbf{\proposal (Ours)} & \cellcolor{highcolor} \correct & \cellcolor{highcolor} \correct & \cellcolor{highcolor} \correct & \cellcolor{highcolor} \correct\\
    \bottomrule
    \end{tabular}
    }
\end{table}

(4) \underline{SWAN}~\cite{hqdl} and \underline{TAG}~\cite{tag} 
design queries that cannot be solely solved by classical SQL but necessitate the involvement of LROs. 
Although they are native to relations, 
they involve only limited LRO types and data domains, resulting in insufficient query diversity and limited single-LRO coverage.
Specifically, SWAN only considers \textsf{Impute}, 
and TAG neglects \textsf{Match} and \textsf{Cluster}.
Moreover, neither of them includes multi-LRO queries, let alone query stratification.

(5) \underline{SemBench}~\cite{sembench} is a benchmark for semantic query processing over multi-modal data,
not native to relations. 
Despite involving multiple types of LROs, SemBench does not explicitly distinguish between single-LRO and multi-LRO workloads.
For single-LRO workloads, 
it fails to cover all type--operand combinations.
Moreover, 
it provides only a limited number of multi-LRO queries without difficulty stratification,
making it difficult to fully reveal system limits.

Last but not least, in the SemBench paper, the evaluation does not dive into the implementation of LROs, 
although the paper notes that ``the implementation of semantic operators matters'' and leaves this as future work.
In contrast, our evaluation in Section~\ref{sec:op-eval} explicitly conducts an in-depth comparison of LROs with different implementations, 
complementing SemBench.

\subsection{\proposal}
\label{sec:bench-new}

\subsubsection{Benchmark Information}
\label{sec:bench-db}

\stitle{Databases.}
\proposal spans 27 real-world databases curated from public benchmarks and repositories: 20 databases from BIRD~\cite{bird}, 5 entity matching databases from the Magellan data repository~\cite{magellandata}, and 2 structured databases from NextiaJD~\cite{nextiajd} and Santos~\cite{santos} data lakes.
These databases span a wide range of domains, including animation, academia, food, books, education, enterprise, crime, software, finance, movies, online shopping, sports, geography, the Olympics, government departments, and city information. 
Moreover, these databases encompass rich semantic attributes that are representative and suitable for LRO workloads.
Table~\ref{tab:benchmark-stats} summarizes the database statistics of \proposal. 
On average, each database contains 11.96 tables, and each table contains 28,064.2 rows and 9.23 columns.

\begin{table}[t]
\centering
\small
\caption{Database Statistics of \proposal.}
\begin{tabular}{lrr}
\toprule
Statistic & Avg. & Max. \\
\midrule
\# tables per database & 11.96 & 80 \\
\# rows per table & 28,064.20 & 1,708,517 \\
\# columns per table & 9.23 & 115 \\
\midrule
\# rows of LRO input tables & 138.85 & 3,208 \\
\# columns of LRO input tables & 10.40 & 77 \\
\bottomrule
\end{tabular}
\label{tab:benchmark-stats}
\end{table}

\begin{new}
\stitle{Query Instances.}
\proposal encompasses 350 query instances, where each query instance can be represented as a $\langle Q, P, R \rangle$ triplet, comprising:
\begin{itemize}
\item \textbf{A natural-language question $Q$} over relational data. By construction, $Q$ cannot be solved by standard SQL alone and requires invoking one or more LROs;
\item \textbf{A procedural plan $P$} that translates $Q$ into an executable program, typically a SQL dialect augmented with LROs. $P$ explicitly specifies the required LRO(s);
\item \textbf{A ground-truth result $R$}, produced by executing $P$ and verified under human supervision to ensure correctness, rather than being directly taken from LLM-generated outputs.
\end{itemize}

Among 350 query instances, 290 of them are single-LRO queries that require only one LRO in their procedural plan, 
and 60 of them are multi-LRO queries demanding multiple LROs to jointly accomplish a complex task.
We present query statistics and detailed examples for single-LRO and multi-LRO queries in Section~\ref{sec:bench-single} and Section~\ref{sec:bench-multi}, respectively.
\end{new}%

\begin{new}
\subsubsection{Annotation Guidelines}
Before data construction, we first specify the annotation guidelines to ensure consistency and reproducibility across annotators.

\stitle{Annotation Schema.}
Each query instance is annotated as a $\langle Q, P, R \rangle$ triplet as stated in Section~\ref{sec:bench-db}:
(i) a natural-language question $Q$;
(ii) an executable procedural plan $P$ (SQL dialect  with explicit LROs);
and (iii) a manually verified ground-truth result $R$.

\stitle{Question ($Q$) Design.} 
First, $Q$ must be constructed based on the schema and contents of a database in our benchmark.
By construction, $Q$ must go beyond the expressive boundary of traditional SQL and thus requires invoking one or more LROs. 
Moreover, $Q$ must be unambiguous under the given databases so that it admits a single deterministic interpretation.

\stitle{Plan ($P$) Construction.}
The plan $P$ must be fully consistent with and equivalent to $Q$, meaning that it completely captures the intent of $Q$ without adding or omitting information.
Moreover, $P$ must be executable in a concrete SQL dialect augmented with LROs that can be directly run in a target system (\eg, LOTUS-TAG~\cite{lotus,tag}, Palimpzest~\cite{palimpzest,abacus}), 
with all invoked LROs and their input arguments explicitly specified.

\stitle{Ground Truth ($R$) Format.}
The ground-truth result $R$ must be a relational table, including degenerate one-cell cases, to support exact-match evaluation. 
Moreover, $R$ must be manually annotated and verified rather than directly taken from LLM-generated outputs, 
to avoid errors caused by LLM hallucinations.

\subsubsection{Annotator Recruitment}
\label{sec:annotator}
We recruit three industry experts (the experts hereinafter) from Alibaba's internal crowdsourcing platform to author and annotate the benchmark.
Each expert is required to: (i) have substantial industry experience in writing and optimizing complex SQL; (ii) understand our LRO taxonomy (Table~\ref{tab:op-tax}) and benchmark principles (Section~\ref{sec:bench-goal}), and strictly follow the prescribed annotation format and guidelines; and (iii) pass an onboarding qualification test on industrial SQL authoring, where we provide ten representative industry-style SQL writing tasks and only experts who pass the test are admitted to the annotation.

Before formal annotation, we train the experts on our benchmark principles and the aforesaid annotation guidelines, 
and familiarize them with executing written procedural plans in the target system through detailed system documentation.
We also provide several representative examples of query instances (later shown in Section~\ref{sec:bench-single} and Section~\ref{sec:bench-multi}) 
to calibrate the expected $\langle Q,P,R\rangle$ format.

\subsubsection{Annotation Workflow}
\label{sec:bench-workflow}
Following the annotation protocol of BIRD~\cite{bird}, our \proposal also adopts a cross-verification workflow as follows.

\stitle{Step 1: Authoring $\langle Q,P,R\rangle$.}
For each assigned query, an expert first explores all databases, including their schema, table contents, and inter-table relationships, to understand the underlying relational structure. 
Then, the expert constructs a $\langle Q, P, R \rangle$ triplet grounded in a database,  ensuring that the query instance is consistent with the database content and captures meaningful real-world semantics.


\stitle{Step 2: Cross-Verification.}
For each query instance authored by one expert, the other two experts independently review it without discussion in the initial pass.
Each reviewer checks three dimensions:
\begin{itemize}
    \item \textbf{Unambiguity:} whether $Q$ admits a single deterministic interpretation under the given databases;
    \item \textbf{Plan validity and sufficiency:} whether $P$ is a faithful and complete operationalization of $Q$;
    \item \textbf{Result correctness:} whether executing $P$ on the input tables yields $R$, and whether $R$ satisfies the requirement of $Q$.
\end{itemize}
A query is accepted only if both reviewers approve it.

\stitle{Step 3: Disagreement Resolution.}
If any reviewer rejects a query, the experts enter an examination-and-revision loop until consensus is reached:
\begin{itemize}
    \item \textbf{Issue identification:} the two reviewers provide concrete reasons, \eg, ambiguity in $Q$, missing step in $P$, incorrect $R$.
    \item \textbf{Revision:} the expert who authors this query  revises $Q$, $P$ or $R$ accordingly. When ambiguity is the root cause, the expert is asked to prioritize revising $Q$ or $P$ to make the intent explicit.
    \item \textbf{Re-verification:} the two reviewers re-check the revised query instance independently.
\end{itemize}
Only query instances that achieve unanimous agreement are included in the final benchmark.

\stitle{Step 4: Final Curation.}
After the experts construct a sufficiently large query corpus (around 500 instances in total), 
we (two of the paper authors) further curate the query corpus to improve diversity and balance.
Specifically, we select query instances to (i) maximize domain and database diversity, (ii) balance LRO types (except \textsf{Filter}), (iii) balance the difficulty of multi-LRO queries, and (iv) remove near-duplicate or overly similar patterns. 
Finally, we curate 290 single-LRO queries and 60 multi-LRO queries.

\emph{Discussion on real-world validity.}
The experts design queries based on real-world industry workloads and BIRD templates~\cite{bird}. 
Some queries adapt BIRD predicates by replacing classical conditions with semantic ones (\eg, ``county is Alameda'' $\rightarrow$ ``county is in Bay Area''), following common practice also used in TAG~\cite{tag}.

\emph{Discussion on table-size constraints.}
Because LLM-generated outputs may contain errors or hallucinations, the ground-truth results for LRO tasks must be carefully verified and annotated by the experts. 
However, validating LRO outputs is labor-intensive, as it often requires cell-by-cell or row-by-row inspection of the entire table, which becomes impractical for large inputs. 
To keep verification tractable while maintaining annotation quality, we ask the experts to constrain the size of input tables during query construction. 
As also illustrated in Table~\ref{tab:benchmark-stats}, the curated LRO instances involve input tables with an average of 138 rows and 10 columns. 
At this scale, the experts can complete manual verification in less than 20 minutes per instance, making high-quality validation feasible.

\end{new}

\begin{figure}[t]
  \centering
  \includegraphics[width=0.9\figwidths]{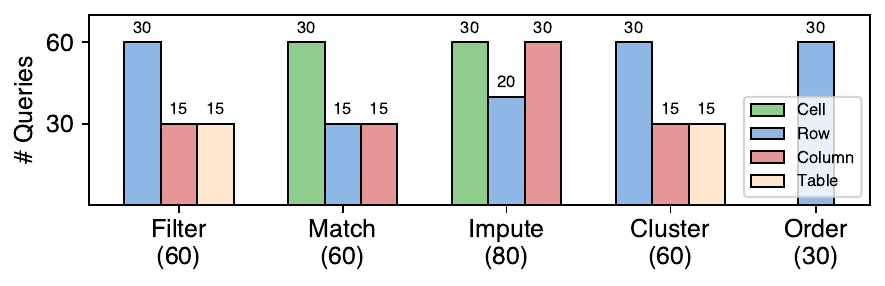}
  \caption{Single-LRO query numbers by LRO types and operands.}
  \label{fig:single-lro-dis}
\end{figure}

\subsubsection{Details of Single-LRO Queries.}
\label{sec:bench-single}
\proposal contains 290 single-LRO queries, each of which contains exactly one LRO in its procedural plan. These queries are designed to evaluate each LRO type under different implementations and design choices (see Section~\ref{sec:op-eval}).
They cover all LRO basic types and fine-grained operands in Table~\ref{tab:op-tax}, including those not yet supported by existing systems, which can help reveal unsupported operator capabilities.
Figure~\ref{fig:single-lro-dis} reports the numbers of the 290 single-LRO queries by LRO types and operands.


\begin{example}[Single-LRO Query]
\label{eg:single-lro}
We introduce an example of a single-LRO query based on the table in Figure~\ref{fig:system-example}.
The query aims to identify and project the columns relevant to the restaurant atmosphere, \ie, a standalone schema linking task, or a semantic column filtering task.
The corresponding plan invokes only one LRO, \texttt{LLM\_FILTER}, 
and the ground truth is the ``\texttt{Description}'' column, annotated by the experts after inspecting all columns.

\smallskip
\begin{mdframed}
\textnormal{\textbf{[Question $Q$]}} Identify and select the columns relevant to the restaurant atmosphere.\\
\textnormal{\textbf{[Plan $P$]}} \textnormal{\texttt{ {\color{darkred}LLM\_FILTER}(Restaurants, ‘column', \\
‘It is related to the atmosphere.');}}\\
\textnormal{\textbf{[Ground Truth $R$]}} A one-column result table containing only the ``\texttt{Description}'' column.
\end{mdframed}
\end{example}

\subsubsection{Details of Multi-LRO Queries.}
\label{sec:bench-multi}
\proposal contains 60 multi-LRO queries, 
each requiring multiple LROs to jointly accomplish a more challenging task.
As in Section~\ref{sec:pre-mt}, 
only a subset of LROs is currently supported for nesting in multi-LRO systems, 
including:
row-wise \textsf{Filter}, 
cell-wise \textsf{Match},
column-wise \textsf{Impute},
row-wise \textsf{Cluster} 
and row-wise \textsf{Order}.
Thus, the experts consider only these composable LROs during construction.

\begin{example}[Multi-LRO Query]
\label{eg:multi-lro}
Consider again the table in Figure~\ref{fig:system-example}.
A multi-LRO query aims to find the restaurant that best suits Asian tastes in the Bay Area.
The corresponding plan sequentially invokes two LROs 
\texttt{LLM\_FILTER} and \texttt{LLM\_ORDER}, 
followed by a classical operator \texttt{LIMIT},
which together return the desired restaurant.
Note that since the table ``\texttt{Restaurants}'' is specified in the main clause, we omit it in the LROs to keep the syntax close to classical SQL.
The ground truth, verified by the experts, is a one-cell table ``\texttt{Alley Wok}''.

\smallskip
\begin{mdframed}
\textnormal{\textbf{[Question $Q$]}} Find the restaurant that best suits Asian tastes in the Bay Area.\\
\textnormal{\textbf{[Plan $P$]}} 
\textnormal{\texttt{
SELECT Name FROM Restaurants\\
WHERE {\color{darkred}LLM\_FILTER}(‘row’, ‘Location is in Bay Area.’) \\
ORDER BY {\color{darkred}LLM\_ORDER}(‘row’, ‘Rank by appeal to Asian tastes from best to worst.')
LIMIT 1;
}}\\
\textnormal{\textbf{[Ground Truth $R$]}} A one-cell  table ``\texttt{Alley Wok}''.
\end{mdframed}
\end{example}





Moreover, to better reveal the capability limits of systems, we ensure the stratification of the multi-LRO query set by scoring these queries along four dimensions:
(i) the number of involved LROs,
(ii) the number of involved tables,
\begin{new}
(iii) the minimum number of operation hops, defined as the number of nodes in the corresponding SQL abstract syntax tree~\cite{DBLP:conf/sigmod/SelingerACLP79}, and
(iv) the complexity of the required knowledge, which is assessed by the experts.
\end{new}%
Specifically, easy-level knowledge pertains to basic common sense, \eg, inferring a continent from a country.
Medium level involves numerical calculation or domain knowledge, 
\eg, obtaining the altitude of race circuits.
Hard level further requires uncommon knowledge, advanced calculation or nuanced reasoning, 
\eg, determining whether a city is within a three-hour drive of New York.

\begin{figure}[t]
  \centering
  \includegraphics[width=0.8\figwidths]{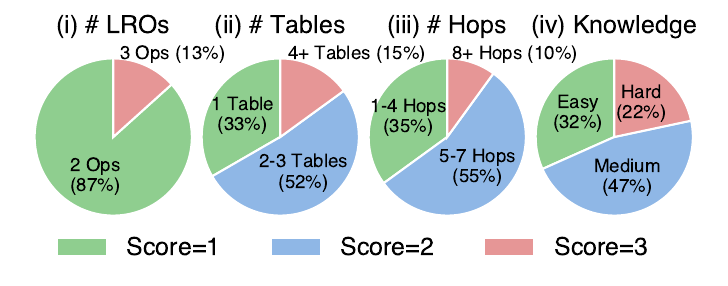}
  \caption{Scoring dimensions of multi-LRO queries.}
  \label{fig:multi-lro-dim}
\end{figure}

\begin{figure}[t]
  \centering
  \includegraphics[width=0.8\figwidths]{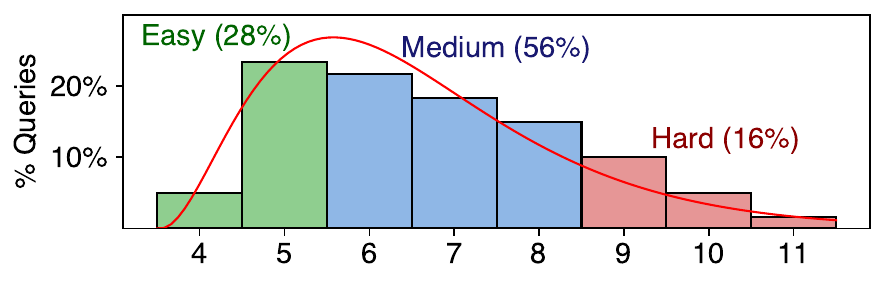}
  \caption{Overall score distribution of multi-LRO queries.}
  \label{fig:multi-lro-score}
\end{figure}

Figures~\ref{fig:multi-lro-dim} and \ref{fig:multi-lro-score} report the distributions of multi-LRO queries, including scores on individual dimensions and overall scores.
Each query is scored on four dimensions, as in Figure~\ref{fig:multi-lro-dim}.
The ``\# LROs'' dimension uses a binary scale: around 87\% of queries contain two LROs (score=1), and the remaining 13\% contain three LROs (score=3)~\footnote{Adding a third LRO substantially increases the combination space and error propagation, so we assign score=3 to distinguish this harder regime.}.
Three-LRO queries are already  challenging for current systems, 
and thus, 
we do not include overly difficult four-LRO queries.
The other three dimensions
are scored from 1 (easy) to 3 (hard): for each of these dimensions, 
around 35\% of queries are easy (score = 1),  50\% are medium (score = 2), 
and the remaining queries are hard (score = 3).
Figure~\ref{fig:multi-lro-score} presents the distribution of overall scores (the sum over all four dimensions) for 60 multi-LRO queries,
which follows a long-tail pattern (the red line).
Easy, medium and hard queries account for 28\%, 56\% and 16\%, respectively.
This is reasonable since the most challenging queries tend to be rare.

\section{Evaluation of Individual LRO}
\label{sec:op-eval}

Based on the taxonomy in Section~\ref{sec:op-tax} and the proposed single-LRO benchmark tasks in Section~\ref{sec:bench-single}, 
we are ready to evaluate each individual LRO under various implementations.
The performance of each individual LRO is evaluated along two dimensions: effectiveness in Section~\ref{sec:op-eval-res}, and scalability in Section~\ref{sec:op-eval-scale}.
Notably, 
we generalize and re-implement all operators (\eg, \texttt{Filter}, \texttt{Match}) and all implementations (\eg, \texttt{LLM-ALL}, \texttt{LLM-ONE}) for a fair single-LRO evaluation.
This is necessary as operators with the same intent are often implemented with different prompts across systems, 
which may impact evaluation fairness.
Moreover, our unified framework allows us to evaluate design choices that have not been explored previously (\eg, applying \texttt{LLM-ALL} to \textsf{Order}, marked with ``---'' in Table~\ref{tab:op-tax}).

In addition, LOTUS-TAG~\cite{lotus,tag} provides an option to enable Chain-of-Thought (CoT) for reasoning. 
We consider this available prompting option for a comprehensive comparison.
By default, we report key results with CoT enabled, 
with full results provided in appendix.

Three state-of-the-art base models are considered: GPT-5~\cite{gpt}, Qwen3-Max~\cite{qwen}, and Claude-Sonnet-4.5~\cite{claude}, representing the latest releases from their respective providers.
We fix the model temperature at 0 and set the context length to the maximum supported by each model.
All results are averaged over three runs.

\subsection{Effectiveness Evaluation} 
\label{sec:op-eval-res}

\subsubsection{{\textsf{Filter}}}
\label{sec:op-eval-select}
\begin{new}
The \textsf{Filter} operator selects rows, columns, or tables based on a semantic condition,
whose result can be represented as a set of selected rows/columns/tables that satisfy the semantic condition.
For each \textsf{Filter} query, we compare the LRO-produced result set with the ground-truth result set, 
and compute precision, recall, and F1 from the overlap between the two sets.
Metrics are computed per query and then averaged over all queries of the same type. 
Table~\ref{tab:op-select} reports these metrics along with LLM costs.
In the table, the best result for each operand task is shown in \textbf{bold} and the second-best is \underline{underlined}; we use this notation throughout the paper.
Unless explicitly stated otherwise, we follow the same evaluation protocol for all subsequent metrics.
\end{new}%
Full results are provided in Table~\ref{app:select} in appendix.

For row-wise filtering, 
\texttt{LLM-ONE} performs strongly across the evaluated models, making it a robust default choice. 
We attribute this to tuple-wise prompting, which provides sufficient per-tuple context and enables fine-grained reasoning for decision making. 
However, for some models (e.g., GPT-5 and Qwen3-Max), the performance gap between \texttt{LLM-ALL} and \texttt{LLM-ONE} becomes small.
Notably, with GPT-5, \texttt{LLM-ALL} even outperforms \texttt{LLM-ONE}.
Meanwhile, \texttt{LLM-ALL} consumes fewer tokens. 
We hypothesize that GPT-5 benefits more from \texttt{LLM-ALL} because its larger context window (400k tokens) and stronger long-context modeling allow it to better leverage many tuples in a single prompt.
In contrast, models with smaller windows and weaker long-context modeling  like Claude-Sonnet-4.5 (200k tokens) may lose effectiveness as the prompt grows.

\begin{table}[t]
\centering
\caption{Key \textsf{Filter} results across base models.}
\label{tab:op-select}
\resizebox{\figwidths}{!}{
\begin{tabular}{c | c | cc | cc | cc} 
\toprule
\multicolumn{1}{c|}{\cellcolor{mygrey}} &
\multicolumn{1}{c|}{\cellcolor{mygrey}} &
\multicolumn{2}{c|}{\cellcolor{mygrey}\textbf{GPT-5}} &
\multicolumn{2}{c|}{\cellcolor{mygrey}\textbf{Qwen3-Max}} &
\multicolumn{2}{c}{\cellcolor{mygrey}\textbf{Claude-Sonnet-4.5}} \\
\multicolumn{1}{c|}{\multirow{-2}{*}{\cellcolor{mygrey}\textbf{Operand}}} &
\multicolumn{1}{c|}{\multirow{-2}{*}{\cellcolor{mygrey}\textbf{Impl}}} &
\cellcolor{mygrey}\textbf{F1} & \cellcolor{mygrey}\textbf{Cost (\$)} &
\cellcolor{mygrey}\textbf{F1} & \cellcolor{mygrey}\textbf{Cost (\$)} &
\cellcolor{mygrey}\textbf{F1} & \cellcolor{mygrey}\textbf{Cost (\$)} \\
\midrule
 
\multirow{2}{*}{Row}& \texttt{LLM-ALL} & \textbf{0.945}  & 0.005 & 0.917 & 0.004 & 0.846 & 0.009  \\
&\texttt{LLM-ONE} 
 &  \underline{0.940} & 0.025 & 0.932  & 0.026 & 0.937 & 0.078\\
 \midrule

\multirow{2}{*}{Column}& \texttt{LLM-ALL} & \textbf{0.891}  & 0.002 & \underline{0.874} & 0.002 & 0.865 & 0.005\\
&\texttt{LLM-ONE}
 &  0.849 & 0.014 & 0.869 & 0.015 & 0.771 & 0.043 \\
 \midrule

\multirow{2}{*}{Table}& \texttt{LLM-ALL} & 0.921  & 0.009 & \underline{0.939} & 0.010 & \textbf{0.943} & 0.024 \\
&\texttt{LLM-ONE} 
 &  0.654 & 0.020 & 0.612  & 0.021 & 0.617 & 0.060 \\
\bottomrule
\end{tabular}
}
\end{table}

For column and table-wise filtering tasks, in Table~\ref{tab:op-select}, 
\texttt{LLM-ALL} consistently performs better than \texttt{LLM-ONE} across different base models.
This is because column and table-wise filtering often requires comparison across candidates to determine which columns or tables are more relevant to the semantic condition.
As \texttt{LLM-ALL} has access to all candidates, it supports such comparisons and enables more accurate identification of the most suitable options.

\insight{1}{
\textbf{\textsf{Filter} Operator Design Choice.} 
For row-wise filtering, \texttt{LLM-ONE} is a strong default across base models, 
while \texttt{LLM-ALL} can be comparable or even better for specific models with strong long-context modeling (\eg, GPT-5) and is more cost-efficient.
For column and table filtering, \texttt{LLM-ALL} is consistently  preferred.
}

\subsubsection{{\textsf{Match}}}
\label{sec:op-eval-match}
The \textsf{Match} operator connects two sets of data elements (cells, rows, or columns) according to a semantic predicate.
Its output is a set of matched pairs (\eg, joined records), 
so we also use \emph{precision}, \emph{recall}, and \emph{F1 score} on prediction against ground truth.
Table~\ref{tab:op-match} reports the key \textsf{Match} evaluation results across models, 
and the full results are provided in Table~\ref{app:match} in appendix.

As shown in Table~\ref{tab:op-match}, 
\texttt{LLM-ALL} achieves the best performance across all cell, row and column-wise \textsf{Match} tasks, 
because \texttt{LLM-ALL} provides a global view of all candidates, 
enabling  relative comparison among similar candidates and reducing inconsistent or conflicting matches that arise from independent pairwise decisions.
For example, in Figure~\ref{fig:example}, consider matching the two tables under the condition ``identify the same individual across Tables~(a) and~(b)''. 
When prompted with all candidates jointly, the LLM can determine that ``Jon R. Moeller'' is the most similar candidate to ``Jon Moeller'' (sharing the same company), and thus the model accepts this pair as a match.
In contrast, \texttt{LLM-ONE} and \texttt{LLM-SEMI} may incorrectly reject this match due to their limited context, which prevents reliable relative comparison, 
and leads to overly loose or overly strict judgment, 
as stated below.


Although \texttt{LLM-ONE} is a mainstream design in prior works~\cite{thalamusdb,palimpzest,abacus,docetl,lotus,tag}, 
it performs poorly with lower F1 score and higher costs on all \textsf{Match} tasks.
This is because each time it considers only a single pair of candidates, 
it may be overly strict (\ie, reject relevant matches) or overly loose (\ie, accept false positives).
For example, when the LLM only sees the pair ``Jon R. Moeller'' and ``Jon Moeller'' in isolation, it may incorrectly determine them as not a match due to the extra middle initial ``R.''.
Moreover, pairwise judgments are not guaranteed to be consistent across comparisons, that is, 
similar candidate pairs may receive different decisions, which further degrades overall performance.

\texttt{LLM-SEMI} generally performs better than \texttt{LLM-ONE} but worse than \texttt{LLM-ALL} as in Table~\ref{tab:op-match}. 
We attribute this to the amount of context it provides: \texttt{LLM-SEMI} uses more context than \texttt{LLM-ONE} but less than \texttt{LLM-ALL}, resulting in intermediate performance.
Notably, on relatively easy row-wise matching tasks with lower ambiguity, \texttt{LLM-SEMI} can approach \texttt{LLM-ALL}.


\begin{table}[t]
\centering
\caption{Key \textsf{Match} results across base models.}
\label{tab:op-match}
\resizebox{\figwidths}{!}{
\begin{tabular}{c | c | cc | cc | cc} 
\toprule
\multicolumn{1}{c|}{\cellcolor{mygrey}} &
\multicolumn{1}{c|}{\cellcolor{mygrey}} &
\multicolumn{2}{c|}{\cellcolor{mygrey}\textbf{GPT-5}} &
\multicolumn{2}{c|}{\cellcolor{mygrey}\textbf{Qwen3-Max}} &
\multicolumn{2}{c}{\cellcolor{mygrey}\textbf{Claude-Sonnet-4.5}} \\
\multicolumn{1}{c|}{\multirow{-2}{*}{\cellcolor{mygrey}\textbf{Operand}}} &
\multicolumn{1}{c|}{\multirow{-2}{*}{\cellcolor{mygrey}\textbf{Impl}}} &
\cellcolor{mygrey}\textbf{F1} & \cellcolor{mygrey}\textbf{Cost (\$)} &
\cellcolor{mygrey}\textbf{F1} & \cellcolor{mygrey}\textbf{Cost (\$)} &
\cellcolor{mygrey}\textbf{F1} & \cellcolor{mygrey}\textbf{Cost (\$)} \\
\midrule
\multirow{3}{*}{Cell}& \texttt{LLM-ALL} & \textbf{0.843} & 0.004 & 0.758 & 0.003 & \underline{0.821} & 0.008  \\
&\texttt{LLM-ONE} & 0.802 & 0.174  & 0.704 & 0.185 & 0.753 & 0.513 \\
&\texttt{LLM-SEMI}  &  0.789 & 0.014 & 0.729 & 0.014 & 0.756 & 0.047 \\
 \midrule

\multirow{3}{*}{Row} & \texttt{LLM-ALL} & \textbf{0.961} & 0.029  & 0.735 & 0.024  & 0.929 & 0.058 \\
&\texttt{LLM-ONE} & 0.829 & 0.338  & 0.862 & 0.335 & 0.797 & 0.946 \\
&\texttt{LLM-SEMI} & 0.934 & 0.085 &  0.922 & 0.079 &  \underline{0.941} & 0.200 \\
 \midrule

\multirow{3}{*}{Column}& \texttt{LLM-ALL} & \textbf{0.746} & 0.003  & \underline{0.735} & 0.003 & 0.726 & 0.008 \\
&\texttt{LLM-ONE} & 0.591 & 0.154 &  0.415 & 0.166 &  0.571 & 0.484 \\
&\texttt{LLM-SEMI} & 0.705 & 0.017 & 0.623 & 0.019 & 0.637 & 0.050 \\
\bottomrule
\end{tabular}
}
\end{table}

\insight{2}{
\textbf{\textsf{Match} Operator Design Choice.}
\texttt{LLM-ALL} generally achieves the best performance on cell, row and column-wise matching tasks, 
a design that has been relatively underexplored in prior works compared with the commonly adopted \texttt{LLM-ONE}.
\texttt{LLM-SEMI} is recommended on row-wise matching tasks with comparable performance.
}

\subsubsection{{\textsf{Impute}}}
\label{sec:op-eval-impute}
The \textsf{Impute} operator augments an incomplete table $R$ by injecting missing cells, columns or rows based on semantics or world knowledge. 
We evaluate the performance via two metrics: 
\emph{Exact Match (EM) ratio}, which measures the ratio of cells with string equivalence, 
and \emph{LLM-as-a-judge score}, where another LLM (different from the LLM in task) judges whether two strings are semantically identical.
Table~\ref{tab:op-impute} presents the key results, 
and Table~\ref{app:impute} in appendix provides the full results.

For the cell-wise imputation tasks in Table~\ref{tab:op-impute}, \texttt{LLM-ALL} and \texttt{LLM-ONE} achieve similar performance, with \texttt{LLM-ALL} showing a slight advantage. For the new-row generation task (row-wise \textsf{Impute}), \texttt{LLM-ALL} significantly outperforms \texttt{LLM-ONE}. 
This suggests that both cell and row-level imputation tasks are highly sensitive to contextual information: when richer context from the relational tables is provided, LLMs are able to generate outputs in a more accurate format.

For the new-column generation tasks (column-wise \textsf{Impute)}, \texttt{LLM-ONE} consistently performs better across different base models.
This is because generating a new column often requires specific fine-grained reasoning and computation, 
for example, generating  a new ``State'' column by inferring from the ``Location'' column.
Each tuple demands distinct external knowledge,
and \texttt{LLM-ONE} provides dedicated reasoning space for each tuple, leading to precise stepwise derivation. 
However, for models with a large maximum context window like GPT-5, 
the performance difference between \texttt{LLM-ONE} and \texttt{LLM-ALL} becomes marginal.


\begin{table}[t]
\centering
\caption{Key \textsf{Impute} results across base models.}
\label{tab:op-impute}
\resizebox{\figwidths}{!}{
\begin{tabular}{c | c | c c | c c | c c}
\toprule
\multicolumn{1}{c|}{\cellcolor{mygrey}} &
\multicolumn{1}{c|}{\cellcolor{mygrey}} &
\multicolumn{2}{c|}{\cellcolor{mygrey}\textbf{GPT-5}} &
\multicolumn{2}{c|}{\cellcolor{mygrey}\textbf{Qwen3-Max}} &
\multicolumn{2}{c}{\cellcolor{mygrey}\textbf{Claude-Sonnet-4.5}} \\
\multicolumn{1}{c|}{\multirow{-2}{*}{\cellcolor{mygrey}\textbf{Operand}}} &
\multicolumn{1}{c|}{\multirow{-2}{*}{\cellcolor{mygrey}\textbf{Impl}}} &
\cellcolor{mygrey}\begin{tabular}[c]{@{}c@{}}\textbf{Judge}\\\textbf{Score}\end{tabular} & \cellcolor{mygrey}\textbf{Cost (\$)} &
\cellcolor{mygrey}\begin{tabular}[c]{@{}c@{}}\textbf{Judge}\\\textbf{Score}\end{tabular} & \cellcolor{mygrey}\textbf{Cost (\$)} &
\cellcolor{mygrey}\begin{tabular}[c]{@{}c@{}}\textbf{Judge}\\\textbf{Score}\end{tabular} & \cellcolor{mygrey}\textbf{Cost (\$)} \\
\midrule

\multirow{2}{*}{Cell}  & \texttt{LLM-ALL} &  \textbf{0.886} & 0.003 &  0.820 & 0.003 &  \underline{0.864} & 0.008 \\
& \texttt{LLM-ONE} &  0.848 & 0.002 & 0.824 & 0.003 & 0.833 & 0.008 \\
\midrule

\multirow{2}{*}{Column} & \texttt{LLM-ALL} & 0.816 & 0.001 & 0.737 & 0.001 & 0.834 & 0.003 \\
& \texttt{LLM-ONE} & \underline{0.838} & 0.005 & 0.793 & 0.006 & \textbf{0.859} & 0.017 \\
\midrule

\multirow{2}{*}{Row} & \texttt{LLM-ALL} & \underline{0.724} & $1.5\times 10^{-4}$ & \textbf{0.751} & $2.1\times 10^{-4}$  & 0.714 & $5.2\times 10^{-4}$  \\
& \texttt{LLM-ONE} & 0.607 & $1.5\times 10^{-4}$ & 0.664 & $1.9\times 10^{-4}$ & 0.561 & $4.8\times 10^{-4}$ \\
\bottomrule
\end{tabular}
}
\end{table}


\insight{3}{
\textbf{\textsf{Impute} Operator Design Choice.}
When generating columns, \texttt{LLM-ONE} is more accurate due to its larger per-tuple reasoning space.
For cell and row-wise imputation, providing sufficient context substantially improves the performance, and thus, 
\texttt{LLM-ALL} is preferred.
}

\subsubsection{{\textsf{Cluster}}}

The \textsf{Cluster} operator divides a set of data elements (rows, columns, or tables) into distinct groups.
We choose two classical clustering measures, namely \emph{adjusted rand index} (ARI)~\cite{hubert1985comparing} and \emph{normalized mutual information} (NMI)~\cite{DBLP:journals/jmlr/StrehlG02}. 
ARI reflects pair-wise clustering accuracy, while NMI calculates the shared information between predicted and ground truth clusters. 
Table~\ref{tab:op-cluster} presents the key \textsf{Cluster} insights with both metrics, 
and Table~\ref{app:cluster} in appendix provides the full results.

In all cases, \texttt{LLM-ALL} outperforms \texttt{LLM-ONE} by a large margin in both ARI and NMI. 
This is simply because \texttt{LLM-ONE} can only process one data element at a time. 
All data elements belonging to the same cluster cannot be perceived simultaneously.
Even worse, during its iteration, errors may propagate. That is, if one element is assigned to a wrong cluster, it may hinder the LLM's understanding of this cluster when processing subsequent elements, further impeding the clustering performance. 
This leads to the following insight:

\insight{4}{
\textbf{\textsf{Cluster} Operator Design Choice.} 
\texttt{LLM-ALL} with global context allows LLMs to formulate a consistent clustering criterion, leading to reliable semantic clustering performance.
Instead, \texttt{LLM-ONE} suffers from criterion drift and error propagation, and thus is not recommended.
}

\begin{table}[t]
\centering
\caption{Key \textsf{Cluster} results across base models.}
\label{tab:op-cluster}
\resizebox{\figwidths}{!}{
\begin{tabular}{c | c | ccc | ccc | ccc} 
\toprule
\multicolumn{1}{c|}{\cellcolor{mygrey}} &
\multicolumn{1}{c|}{\cellcolor{mygrey}} &
\multicolumn{3}{c|}{\cellcolor{mygrey}\textbf{GPT-5}} &
\multicolumn{3}{c|}{\cellcolor{mygrey}\textbf{Qwen3-Max}} &
\multicolumn{3}{c}{\cellcolor{mygrey}\textbf{Claude-Sonnet-4.5}} \\
\multicolumn{1}{c|}{\multirow{-2}{*}{\cellcolor{mygrey}\textbf{Operand}}} &
\multicolumn{1}{c|}{\multirow{-2}{*}{\cellcolor{mygrey}\textbf{Impl}}} &
\cellcolor{mygrey}\textbf{ARI} & \cellcolor{mygrey}\textbf{NMI} & \cellcolor{mygrey}\textbf{Cost (\$)} &
\cellcolor{mygrey}\textbf{ARI} & \cellcolor{mygrey}\textbf{NMI} & \cellcolor{mygrey}\textbf{Cost (\$)} &
\cellcolor{mygrey}\textbf{ARI} & \cellcolor{mygrey}\textbf{NMI} & \cellcolor{mygrey}\textbf{Cost (\$)} \\
\midrule

\multirow{2}{*}{Row}  & \texttt{LLM-ALL} & \textbf{0.723} & \textbf{0.801} & 0.008  & \underline{0.720} & \underline{0.778} & 0.006  & 0.675 & 0.763 & 0.016 \\
& \texttt{LLM-ONE} & 0.694 & 0.759 & 0.029  & 0.658 & 0.727 & 0.026  & 0.696 & 0.756 & 0.073 \\
\midrule

\multirow{2}{*}{Column} & \texttt{LLM-ALL} & \underline{0.812} & \underline{0.877} & 0.010  & 0.751 & 0.835 & 0.008  & \textbf{0.848} & \textbf{0.897} & 0.018 \\
& \texttt{LLM-ONE} & 0.604 & 0.721 & 0.028  & 0.493 & 0.641 & 0.027  & 0.610 & 0.726 & 0.077 \\
\midrule

\multirow{2}{*}{Table} & \texttt{LLM-ALL} & \textbf{0.721} & \textbf{0.837} & 0.015  & 0.627 & 0.783 & 0.015  & \underline{0.716} & \underline{0.830} & 0.035 \\
& \texttt{LLM-ONE} & 0.668 & 0.801 & 0.023 & 0.630 & 0.780 & 0.023 & 0.647 & 0.788 & 0.059 \\
\bottomrule
\end{tabular}
}

\end{table}

\subsubsection{{\textsf{Order}}}
\label{sec:op-eval-order}
The \textsf{Order} operator ranks a set of records (e.g., rows or items) according to a semantic criterion (e.g., "popularity" or "relevance"). 
We utilize \emph{Top-$k$ hit rate}~\cite{DBLP:conf/recsys/CremonesiKT10}, denoted as HR@$k$, and \emph{Kendall's $\tau$}~\cite{kendall1938new}, denoted as $\tau$. 
HR@$k$ measures how many ground-truth top-$k$ records are retrieved in the prediction, 
and $\tau$ further measures ranking consistency on the hit tuples, 
\ie, it evaluates whether their relative order matches the ground-truth ranking order.
Table~\ref{tab:op-order} presents the \textsf{Order} results and Table~\ref{app:order} in appendix further provides full results.

As shown, \texttt{LLM-ALL} achieves the highest HR@$k$ with Qwen3-Max and the highest $\tau$ with GPT-5, 
demonstrating its superior performance. 
\texttt{LLM-PAIR} and \texttt{LLM-SORT} perform moderately but suffer from two critical issues.
First, they lack global perception, 
thus difficult to maintain coherent semantic preferences across comparisons; 
second, more severely, they are vulnerable to \emph{inconsistent pairwise judgments}. Due to the stochastic nature of LLM outputs, contradictions such as $A > B$, $B > C$, yet $C > A$ frequently occur, violating transitivity and leading to logically invalid rankings. These inconsistencies propagate through sorting algorithms (\eg, quicksort), 
amplifying errors and undermining overall correctness.
\texttt{LLM-SCORE} performs worst due to low score discriminative power. 
LLMs often assign nearly identical values (\eg, all 80 or 90) under vague criteria, 
which collapses the ranking into arbitrary tie-breaking.

\begin{table}[t]
\centering
\caption{Key row-wise \textsf{Order} results across base models.}
\label{tab:op-order}
\resizebox{\figwidths}{!}{
\begin{tabular}{ c | ccc | ccc | ccc} 
\toprule
 \multicolumn{1}{c|}{\cellcolor{mygrey}} &
\multicolumn{3}{c|}{\cellcolor{mygrey}\textbf{GPT-5}} &
\multicolumn{3}{c|}{\cellcolor{mygrey}\textbf{Qwen3-Max}} &
\multicolumn{3}{c}{\cellcolor{mygrey}\textbf{Claude-Sonnet-4.5}} \\
\multicolumn{1}{c|}{\multirow{-2}{*}{\cellcolor{mygrey}\textbf{Impl}}}  &
\cellcolor{mygrey}\textbf{HR@$k$} & \cellcolor{mygrey}\textbf{$\tau$} & \cellcolor{mygrey}\textbf{Cost (\$)} &
\cellcolor{mygrey}\textbf{HR@$k$} & \cellcolor{mygrey}\textbf{$\tau$} & \cellcolor{mygrey}\textbf{Cost (\$)} &
\cellcolor{mygrey}\textbf{HR@$k$} & \cellcolor{mygrey}\textbf{$\tau$} & \cellcolor{mygrey}\textbf{Cost (\$)} \\
\midrule

\texttt{LLM-ALL}   & 0.941 & \textbf{0.937} & 0.002 & \textbf{0.952} & 0.884 &  0.002  & \underline{0.949} & 0.865 & 0.006  \\
\texttt{LLM-PAIR}  & 0.929 & 0.907 & 0.182 & 0.921 & 0.892 & 0.208 & 0.910 & \underline{0.914} & 0.568 \\ 
\texttt{LLM-SORT}  & 0.896 & 0.881 & 0.021 & 0.941 & 0.884 & 0.033 & 0.899 & 0.871 & 0.083 \\
\texttt{LLM-SCORE} & 0.776 & 0.658 & 0.011 & 0.805 & 0.588 & 0.014 & 0.798 & 0.706 & 0.039 \\
\bottomrule
\end{tabular}
}
\end{table}

\insight{5}{
\textbf{\textsf{Order} Operator Design Choice.}
\texttt{LLM-ALL} is strongly recommended, as it delivers the best performance at the lowest cost. 
Both \texttt{LLM-PAIR} and \texttt{LLM-SORT} methods are vulnerable to inconsistent judgments.
\texttt{LLM-SCORE}, which lacks a reasonable sense of scores, is not recommended.
}

\subsubsection{Best Practices}
\label{sec:best-practice}
We summarize the insights from \iref{1} to \iref{5} into best practices in Table~\ref{tab:best-practices},
where ``---'' denotes the rare combinations of LROs and operands, 
as stated in Section~\ref{sec:op-tax}.
We highlight different design choices using colored boxes (with the same color coding as Table~\ref{tab:op-tax}).

\nop{
\begin{table}[t]
\centering
\caption{Ablation study on chain-of-thought (CoT)  for Cell-wise \textsf{Impute}. }
\label{tab:cot-ex}
\resizebox{\figwidths}{!}{
\begin{tabular}{ c | c c | c |c  | c}
\toprule
\multicolumn{1}{c|}{\cellcolor{mygrey}} &
\multicolumn{2}{c|}{\cellcolor{mygrey}} &
\cellcolor{mygrey}\textbf{GPT-5} &
\cellcolor{mygrey}\textbf{Qwen3-Max} &
\cellcolor{mygrey}\textbf{Claude-Sonnet-4.5} \\
\multicolumn{1}{c|}{\multirow{-2}{*}{\cellcolor{mygrey}\textbf{Impl}}} &
\multicolumn{1}{c}{\multirow{-2}{*}{\cellcolor{mygrey}\textbf{CoT}}} &
\multicolumn{1}{c|}{\multirow{-2}{*}{\cellcolor{mygrey}\textbf{EX}}} &
\cellcolor{mygrey}\textbf{Judge Score} &
\cellcolor{mygrey}\textbf{Judge Score} &
\cellcolor{mygrey}\textbf{Judge Score} \\
\midrule

\multirow{4}{*}{\texttt{LLM-ONE}} & \wrong  & \wrong & 0.681 & 0.634  & 0.643  \\
 & \correct & \wrong & 0.686 & 0.673 & 0.654  \\
 & \wrong & \correct & 0.836 & 0.746  & 0.792 \\
& \correct & \correct & 0.848 & 0.824 & 0.833  \\
\bottomrule
\end{tabular}
}%
\end{table}
}

\begin{table}
\centering
\caption{Best practices of LRO designs.}
\label{tab:best-practices}
\resizebox{0.8\figwidths}{!}{
\begin{tabular}{r| cccc}
\toprule
\rowcolor{mygrey} \textbf{LRO} & \textbf{Cell} & \textbf{Row} & \textbf{Column} & \textbf{Table} \\
\midrule
\textsf{Filter} & --- & \LLMONE & \LLMALL & \LLMALL \\
\textsf{Match}  & \LLMALL &  \LLMSEMI & \LLMALL  & --- \\
\textsf{Impute} & \LLMALL  & \LLMONE & \LLMONE & --- \\
\textsf{Cluster} & --- & \LLMALL  & \LLMALL & \LLMALL\\
\textsf{Order} & --- & \LLMALL  & --- & --- \\
\bottomrule
\end{tabular}
}
\end{table}

\begin{figure}[t]
\centering
\begin{subfigure}{0.5\linewidth}
  \centering
  \includegraphics[width=\linewidth]{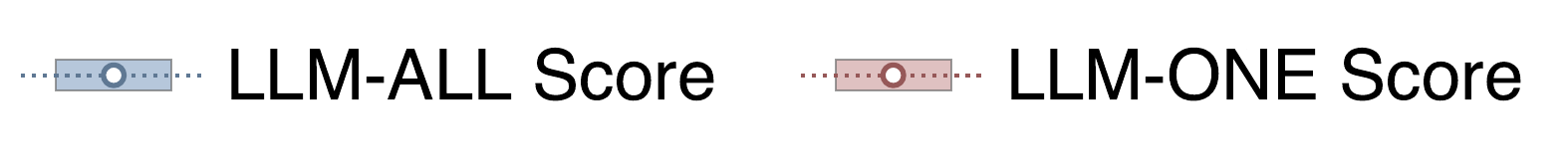}
  \includegraphics[width=\linewidth]{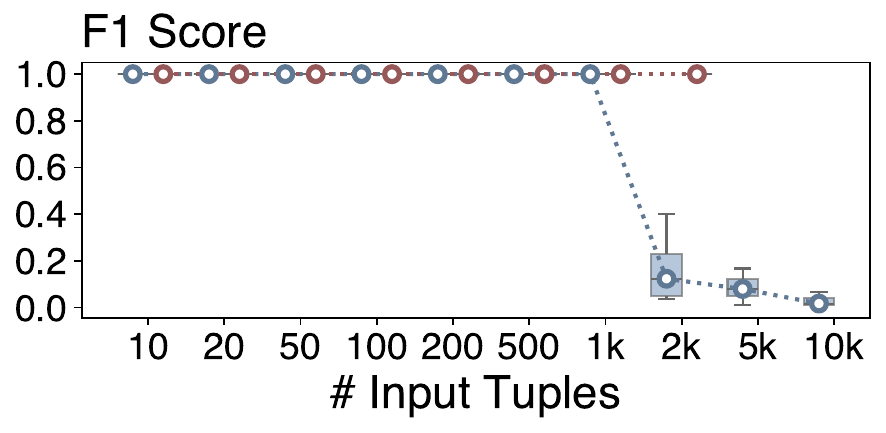}
\end{subfigure}\hspace{-0.5em}
\begin{subfigure}{0.5\linewidth}
  \centering
  \includegraphics[width=\linewidth]{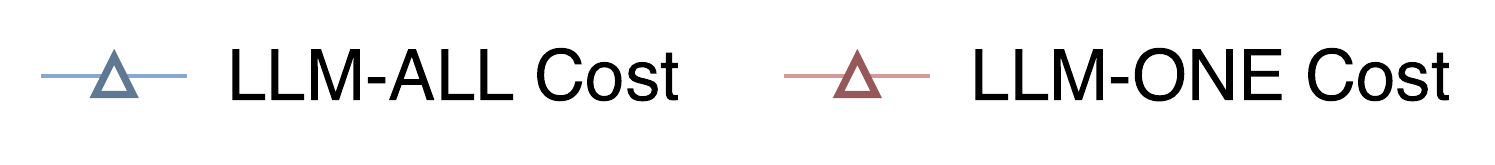}
  \includegraphics[width=\linewidth]{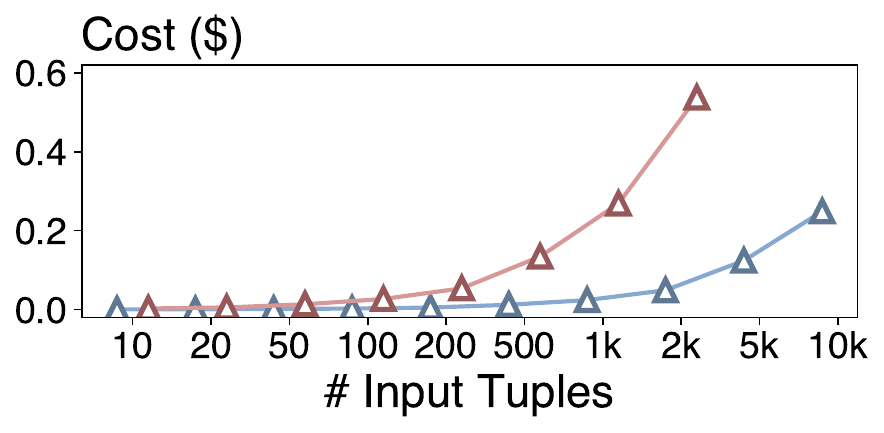}
\end{subfigure}
\caption{Effectiveness and cost of \textsf{Filter} at scale.}
\label{exp:scale-select}
\end{figure}

\begin{figure}[h]
\centering
\begin{subfigure}{0.5\linewidth}
  \centering
  \includegraphics[width=\linewidth]{exps/scale/score_legend.pdf}

  \includegraphics[width=\linewidth]{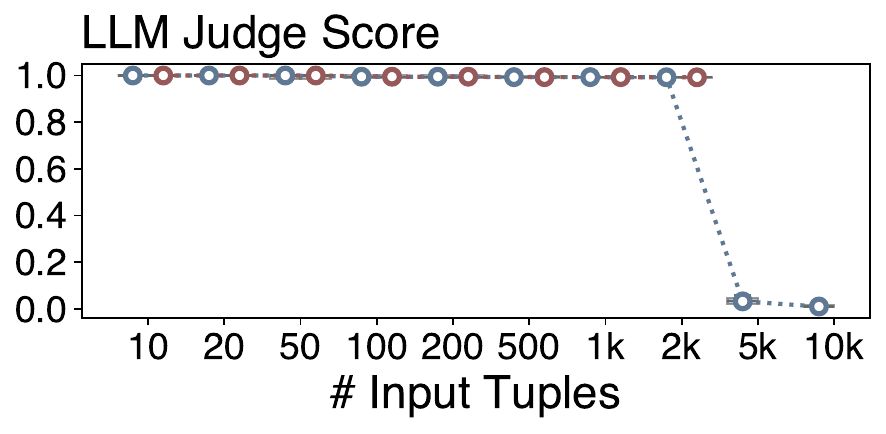}
\end{subfigure}\hspace{-0.5em}
\begin{subfigure}{0.5\linewidth}
  \centering
  \includegraphics[width=\linewidth]{exps/scale/cost_legend.pdf}
  \includegraphics[width=\linewidth]{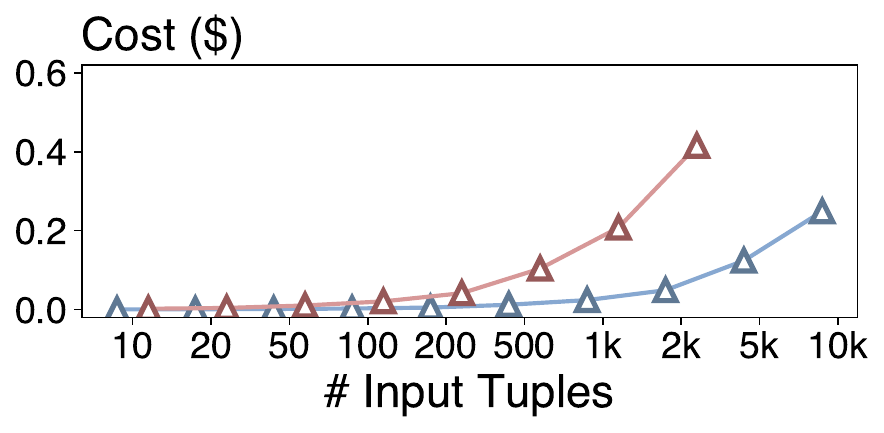}
\end{subfigure}
\caption{Effectiveness and cost of \textsf{Impute} at scale.}
\label{exp:scale-impute}
\end{figure}

\subsection{Scalability Evaluation}
\label{sec:op-eval-scale}

While effectiveness is critical, real-world LRO query tasks may often involve large-scale inputs, 
making scalability a key practical concern.
We conduct scalability experiments on the two most prevalent tasks for large-scale data: row-wise \textsf{Filter} for semantic filtering  in Figure~\ref{exp:scale-select},
and column-wise \textsf{Impute} for  augmentation in Figure~\ref{exp:scale-impute}.
Similar to Sections~\ref{sec:op-eval-select} and \ref{sec:op-eval-impute}, the methods
\texttt{LLM-ALL} and \texttt{LLM-ONE} are considered.
All experiments run on a ``european\_football\_2'' database from BIRD~\cite{bird} with more than 10k records.
Experiments are repeated 10 times under GPT-5 to capture variations (with results shown as box plots), 
and each run has a 30-minute timeout.
As ground truth for semantic filtering or imputation over large tables is hard to obtain, 
we design two queries whose ground truth can be determined equivalently by deterministic rules:
(i) \emph{``Identify the players born after the fall of the Berlin Wall''} for \textsf{Filter}, 
and (ii) \emph{``Determine the zodiac sign for each player based on their date of birth''} for \textsf{Impute}.

The \texttt{LLM-ALL} method degrades rapidly in accuracy and stability,
and fails at relatively small scales (2k--10k tuples) despite input token counts remaining much below the 400k token limit of GPT-5.
At this point, outputs become malformed or syntactically invalid, indicating a breakdown in structured generation. 
This reveals a critical gap: 
\emph{the effective context boundary for reliable reasoning is far smaller than the theoretical maximum}. 
We hypothesize that long prompts overload the LLM’s attention modules, 
diverting attention from key conditions and disrupting coherent response formation,
even when technically ``within limits''.


\insight{6}{
\textbf{Limited Effective Context Window of \texttt{LLM-ALL}.}
Although within token limit, LLMs may fail to maintain structured generation under large-scale inputs due to attention dilution.
}

In contrast, \texttt{LLM-ONE}  
performs fine-grained reasoning and consistently maintains high quality (\eg, F1$=1.0$ on \textsf{Filter}) until it hits the timeout limit, but it incurs higher inference costs than \texttt{LLM-ALL}.

As discussed in Section~\ref{sec:op-select}, \texttt{LLM-ALL} and \texttt{LLM-ONE} represent two extreme design points: 
\texttt{LLM-ALL} is cost-optimal, whereas \texttt{LLM-ONE} is quality-optimal.
However, Figures~\ref{exp:scale-select} and \ref{exp:scale-impute} show that \texttt{LLM-ALL} remains high-quality under GPT-5 when batching fewer than 100 tuples (around 2,000 input tokens) per request.
This suggests that: there exists a hybrid implementation between the two design extremes (\ie, \texttt{LLM-ALL} \vs  \texttt{LLM-ONE}) to make the best trade-off between two objectives (\ie, quality \vs cost). For example, prompting GPT-5 with a small batch of 50 tuples (around 1,000 tokens) per request can achieve optimal quality at near-optimal cost. 
For other base models, we can characterize an LRO design by a model-specific quality--cost curve, with token number on the x-axis and result quality on the y-axis, to find the best balance point. Such detailed explorations are left to our future work.

\insight{7}{
\textbf{LRO Design Trade-off.}
LRO design entails a trade-off that can be cast as optimizing over a model-specific quality--cost curve. Smart batching strategies and fine-tuned LRO implementations integrating \texttt{LLM-ALL} and \texttt{LLM-ONE} are expected to make best trade-offs between quality and cost.
}%




\section{Evaluation of Multi-LRO Systems}
\label{sec:e2e-eval}

\begin{table*}[t]
\centering
\small
\caption{LRO correspondences and benchmark query coverage of existing multi-LRO systems.}
\label{tab:correspond}
\resizebox{0.85\textwidth}{!}{
\begin{tabular}{r|ccccc|c} 
\toprule
\rowcolor{mygrey}
 \textbf{System} &  \textbf{\textsf{Filter}} &  \textbf{\textsf{Match}} &  \textbf{\textsf{Impute}}  
& \textbf{\textsf{Cluster}} & \textbf{\textsf{Order}} 
&  \textbf{Query Coverage}\\
\midrule
Binder & \texttt{f\_col} + \underline{\texttt{WHERE}}
    & \wrong 
    & \texttt{f\_col}
    & \texttt{f\_col} + \underline{\texttt{GROUPBY}}
    & \texttt{f\_col} + \underline{\texttt{ORDERBY}} 
    & 33.3\% \\
SUQL & \texttt{Answer} + \underline{\texttt{WHERE}}
    & \wrong 
    & \texttt{Answer} 
    & \texttt{Answer} + \underline{\texttt{GROUPBY}}
    & \texttt{Answer} + \underline{\texttt{ORDERBY}} 
    & 75.0\%\\
Aryn-Sycamore &  \texttt{Filter}
    & \wrong
    & \texttt{Map}
    & \texttt{Map} + \underline{\texttt{GROUPBY}}
    & \texttt{Map} + \underline{\texttt{ORDER}}
    & 33.3\%\\
HQDL & \texttt{Generate} +\underline{\texttt{WHERE}} 
    & \wrong 
    & \texttt{Generate} 
    & \texttt{Generate}  + \underline{\texttt{GROUPBY}} 
    & \texttt{Generate}  + \underline{\texttt{ORDERBY}} 
    & 75.0\%\\
ThalamusDB & \texttt{NLFILTER} 
    & \texttt{NLJOIN}
    & \wrong
    & \wrong
    & \wrong
    & 41.7\%\\
BlendSQL & \texttt{LLMValidate}  
    & \texttt{LLMJoin} 
    & \texttt{LLMMap} 
    & \texttt{LLMMap} + \underline{\texttt{GROUPBY}} 
    & \texttt{LLMMap} + \underline{\texttt{ORDERBY}}
    & 100.0\% \\
Palimpzest & \texttt{Filter} 
    & \texttt{Join} 
    & \texttt{Map} 
    & \texttt{Map} + \underline{\texttt{GROUPBY}} 
    & \texttt{Map} + \underline{\texttt{ORDERBY}} 
    & 100.0\% \\
DocETL & \texttt{Filter} 
    & \texttt{Merge} 
    & \texttt{Map} 
    & \texttt{Map} + \underline{\texttt{GROUPBY}} 
    & \texttt{Map} + \underline{\texttt{ORDERBY}} 
    & 100.0\% \\
LOTUS-TAG & \texttt{sem\_filter} 
    & \texttt{sem\_join} 
    & \texttt{sem\_map} 
    & \texttt{sem\_cluster\_by} 
    & \texttt{sem\_topk}  
    & 100.0\% \\ \hline
Our Best Practices  & \textsf{Filter} 
    & \texttt{Match} 
    & \texttt{Impute} 
    & \texttt{Cluster} 
    & \texttt{Order}  
    & 100.0\% \\
\bottomrule
\end{tabular}}
\end{table*}

\begin{table*}[t]
    \centering
    \caption{Performance of multi-LRO systems on multi-LRO benchmark.}
    \label{tab:e2e-res}
    \resizebox{\textwidth}{!}{
    \begin{tabular}{c|c|ccc|ccccccc}
    \toprule
    \multicolumn{2}{c|}{\cellcolor{mygrey}} 
    & \multicolumn{3}{c|}{ \cellcolor{mygrey}\textbf{LLM-Planning Systems}} & \multicolumn{7}{c}{ \cellcolor{mygrey} \textbf{Manual-Planning Systems}}   \\
    
    \multicolumn{2}{c|}{\cellcolor{mygrey}\textbf{Metric}} 
    & \cellcolor{mygrey} & \cellcolor{mygrey} & \cellcolor{mygrey} & \cellcolor{mygrey} & \cellcolor{mygrey} & \cellcolor{mygrey} & \cellcolor{mygrey} & \cellcolor{mygrey} & \cellcolor{mygrey} & \cellcolor{mygrey}\\ 
    \multicolumn{2}{c|}{\cellcolor{mygrey}} 
    &  \multirow{-2}{*}{ \cellcolor{mygrey} Binder} & \multirow{-2}{*}{ \cellcolor{mygrey} SUQL} & \multirow{-2}{*}{\cellcolor{mygrey} \begin{tabular}[c]{@{}c@{}}Aryn-\\[-0.1em] Sycamore\end{tabular}}  & \multirow{-2}{*}{\cellcolor{mygrey} HQDL} 
    & \multirow{-2}{*}{\cellcolor{mygrey} ThalamusDB}
    & \multirow{-2}{*}{\cellcolor{mygrey} BlendSQL}
    & \multirow{-2}{*}{\cellcolor{mygrey} Palimpzest}  
    & \multirow{-2}{*}{\cellcolor{mygrey} DocETL}
    & \multirow{-2}{*}{\cellcolor{mygrey} LOTUS-TAG} 
    & \multirow{-2}{*}{\cellcolor{mygrey} \begin{tabular}[c]{@{}c@{}} \bf Our Best\\[-0.1em] \bf Practices \end{tabular}}\\
    \midrule 
    
    \multirow{2}{*}{\textbf{GPT-5}}
    &  Accuracy & 3.33\% & 13.33\% & 0\%  & 30.00\% & 33.33\% &  55.00\% &  43.33\% & 13.33\% & 50.00\% & \textbf{86.67\%} \\ 
    & Cost (\$) & 0.44 & 0.02 &  0.08 & 1.13 & 0.12 &  0.12 & 0.46 & 0.78 &  0.40 & 0.17\\ 
    \midrule
    \multirow{2}{*}{\textbf{Qwen3-Max}} & Accuracy & 3.33\%  & 6.67\%  & 0\%  & 31.67\%  &  15.00\% &  31.67\% & 70.00\%  & 33.33\% & 38.33\% & \textbf{71.67\%} \\ 
    & Cost (\$) & 0.24  & 0.01 & 0.36 & 1.24 & 0.03 & 0.03 & 0.32 & 0.41 &  0.15 & 0.06\\ 
    \midrule
    \multirow{2}{*}{\makecell[c]{\bf Claude-\\ 
    \bf Sonnet-4.5}} & Accuracy & 0\%  &  6.67\% & 0\% & 33.33\% &  8.33\% &  35.00\% &  66.67\% & 40.00\% & 38.33\% & \textbf{76.67\%} \\ 
    & Cost (\$) & 0.91 & 0.03 & 0.05 &  2.42 &  0.16 &  0.10 &  0.80 & 1.45 & 0.35 & 0.24\\ 
    \midrule
    \addlinespace[0.5ex]
    \midrule
    \multirow{3}{*}{\makecell[c]{\bf Stratified  \\ \bf Accuracy \\ \bf under GPT-5}}
    & Easy (n=17) &  11.76\% & 23.53\% & 0\% & 35.29\% &  35.29\%  & 70.59\%  & 52.94\% & 5.88\%  & 64.71\% & \textbf{100.00\%} \\
    & Medium (n=33) &  0\%  & 12.12\% & 0\% & 30.30\%  & 33.33\%  &  54.55\%  & 42.42\% & 18.18\%  & 51.52\% & \textbf{81.82\%} \\
    & Hard (n=10) & 0\% & 0\% & 0\% & 20.00\% &  10.00\%  & 30.00\%  & 30.00\%  &  10.00\% & 30.00\% & \textbf{80.00\%} \\ 
    \bottomrule
    \end{tabular}}
\end{table*}

While Section \ref{sec:op-eval} discusses the optimal design choices for each individual LRO type, designing a multi-LRO solution goes beyond optimizing each LRO in isolation, requiring a deeper understanding of how to integrate multiple LROs into a coherent end-to-end system.
Therefore, in this section, we further explore 
how to orchestrate our operator-level best practices,
and compare them against existing multi-LRO systems.
Section~\ref{sec:e2e-setup} details the evaluation setups, 
and Section~\ref{sec:e2e-res} presents the evaluation results.


\subsection{Evaluation Setups}
\label{sec:e2e-setup}

\subsubsection{Query Set}
As in Sections~\ref{sec:pre-mt} and \ref{sec:bench-multi}, only a subset of LROs are commonly supported for nesting in current systems: row-wise \textsf{Filter}, cell-wise \textsf{Match}, column-wise \textsf{Impute}, row-wise \textsf{Cluster}, and row-wise \textsf{Order}.
Accordingly, we use the challenging multi-LRO queries from our \proposal, 
which contains 60 expert-curated queries composed of these five LROs and stratified by complexity.%

\subsubsection{Baselines}
\label{sec:e2e-baselines}
Our baselines are the existing multi-LRO systems introduced in Section~\ref{sec:pre-mt}, 
which provide an LRO set seamlessly integrated with SQL, featuring some (maybe all) of the five composable LROs above.
Some systems orchestrate LROs and SQL operators into a query plan by automated LLM-based planning, \ie, LLM-planning systems, 
such as Binder~\cite{binder}, SUQL~\cite{suql} and Aryn-Sycamore~\cite{aryn}.
They typically take natural-language questions as input, and internally translate them into a procedural plan.
Other systems~\cite{hqdl,thalamusdb,blendsql,palimpzest,abacus,docetl,lotus,tag} rely on manually annotated query plans (\eg, Python code or a SQL-like dialect), \ie, manual-planning systems.
They can complete a task only by executing an executable procedural plan written by the experts.

Fortunately, our benchmark in Section~\ref{sec:bench-multi} provides query tasks annotated with both natural-language questions and procedural plans. 
Therefore, we provide natural-language questions for automated LLM-planning systems. 
For each manual-planning system, we invite the experts (the three industry experts in Section~\ref{sec:annotator}) to recast the benchmark procedural plans into the system’s required syntax and ensure correctness.
All systems are reproduced from their open-source repositories. 

Notably, we further construct a manual-planning baseline by composing our single-LRO empirical best practices with their optimal implementations as in Section~\ref{sec:best-practice} and Table~\ref{tab:best-practices}.
Specifically, we use \texttt{LLM-ONE} for row-wise \textsf{Filter} and column-wise \textsf{Impute}; 
we use \texttt{LLM-ALL} for cell-wise \textsf{Match}, row-wise \textsf{Cluster} and  row-wise  \textsf{Order}.

\stitle{Support of LROs.}
Not all baseline systems explicitly support all five LROs in our benchmark.
Table~\ref{tab:correspond} illustrates their  
LRO correspondences and query coverage (\ie, how many queries fall within their scope).
For LLM-planning systems, we preserve their native planning strategies,
allowing free composition of LROs and SQL operators.
Note that Binder and Aryn-Sycamore only support a single table as input, resulting in coverage of only one-third of the queries.
For manual-planning systems, each query $Q$ requiring LROs $O$ is manually planned using the following rules to maximize attainable system performance: 

$\bullet$  If $O$ is explicitly implemented (often under a similar name), we directly invoke the native LRO (marked by its name in Table~\ref{tab:correspond}). 
For instance, \texttt{sem\_select} in LOTUS-TAG corresponds to \textsf{Filter}.

$\bullet$ Otherwise, 
we realize  $O$ indirectly by composing other available LROs and classical SQL operators (marked in the form of ``LRO + \underline{SQL operator}'').
For example, SUQL combines its \texttt{Answer} LRO with classical \texttt{WHERE} clause to perform semantic filtering.

$\bullet$ In the worst case, the system is completely incapable of realizing the logic of $O$, 
thus fails on query $Q$ (marked as a red cross).

\nop{
For all handwritten-planning methods, experts adopt similar strategies to annotate query plans, 
aiming to address as many queries as possible.
Explicitly supported operators are directly invoked by experts. 
Instead, for implicitly supported operators, 
their functional equivalence is achieved through semantic-to-SQL translation or operator composition. 
}%

\subsubsection{Metrics}
\label{sec:e2e-metrics}
Since multi-LRO systems compose LROs (as in Definition~\ref{def:lro}) with classical SQL operators, they produce a table-valued result for a given query. 
Evaluating such results requires comparing two tables rather than two scalar values or ranked lists, which calls for a stricter and more structured evaluation criterion.
We therefore adopt \emph{Exact Match} (EM) between the predicted table (\ie, the output table of the system) and the ground-truth table as our evaluation criterion. 
Specifically, a query is counted as correctly answered if and only if the two tables are identical.
We report the EM accuracy as our primary evaluation metric, defined as the fraction of queries for which the predicted table exactly matches the ground-truth table, out of the 60 multi-LRO queries in our benchmark. 


\subsubsection{Configurations}
Consistent with Section~\ref{sec:op-eval}, we employ 
GPT-5~\cite{gpt}, Qwen3-Max~\cite{qwen}, and Claude-Sonnet-4.5~\cite{claude}.
We set the temperature to 0 and use the maximum context length supported by each model.
For systems that support parallel LLM invocations, we set the parallelism to 10, and apply a 30-minute timeout per query.

\subsection{Evaluation Results}
\label{sec:e2e-res}

\subsubsection{Overall Performance Analysis}
\label{sec:e2e-res-pa}

Table~\ref{tab:e2e-res} reports the overall performance of multi-LRO baselines on our multi-LRO query set.
Overall, LLM-planning systems perform poorly (all below 15\% accuracy), substantially lagging behind manual-planning systems.
This gap is largely due to failures in the initial planning step, which yield flawed query plans and propagate errors throughout the pipeline.
Even with expert-annotated plans, however, most manual-planning systems (except Palimpzest) still achieve limited accuracy (below 55\%).
In contrast, our best-practice approach achieves the best results across all three base models, peaking at 86.67\% with GPT-5.

Table~\ref{tab:e2e-res} also reports GPT-5 accuracy by difficulty, with 17 easy, 33 medium, and 10 hard queries. Overall, current systems struggle on hard queries. 
Notably, no existing LLM-planning system solves any hard query (all 0\% accuracy), indicating substantial room for improvement in automatic LRO planning. 
Even manual-planning systems, despite using annotated plans, 
solve at most 3 of the 10 hard queries and about half of the medium queries.
In contrast, our best-practice baseline achieves 100\%, 81.82\%, and 80\% accuracy on easy, medium, and hard queries, respectively, 
 due to its full LRO coverage and near-optimal LRO implementations.

\insight{8}{\textbf{Early-Stage and Room for Improvement.}
Current LLM-planning systems perform poorly on our benchmark ($<$ 15\% accuracy) and fail on complex queries, 
as unreliable initial plans derail downstream steps.
Even with expert-annotated plans, most manual-planning systems (except Palimpzest) remain below 55\%.
Overall, automatic LRO planning remains early-stage and multi-LRO systems still have substantial room for improvement.
}%

\insight{9}{\textbf{Best Practices as Guidelines.}
Our best-practice baseline consistently performs best and solves up to 80\% of hard queries, 
suggesting it may serve as a useful guideline for future systems.
}


\subsubsection{Performance \vs LRO Coverage and Design}
\label{sec:e2e-lro-design}
We next analyze how LRO coverage and LRO design affect the overall performance.
Consistent with Table~\ref{tab:correspond}, 
systems that support broader coverage of our LRO taxonomy and cover more queries generally achieve higher accuracy.
For LLM-planning systems, Binder and Aryn-Sycamore only support single-table queries, rendering two-thirds multi-table queries out of scope, and thus perform poorly.
SUQL covers 75\% of the queries and achieves correspondingly higher accuracy.
For manual-planning systems, BlendSQL, Palimpzest, and LOTUS-TAG cover all queries in our benchmark and therefore exceed 50\% peak accuracy. 
DocETL also covers all queries but peaks at around 40\%. 
In contrast, HQDL and ThalamusDB respectively cover only 75\% and 41.7\% of the queries, limiting their peak accuracy to around 30\%.


How LROs are implemented also matters.
Systems whose LRO designs more closely follow our best-practice implementations tend to perform better, such as Palimpzest and LOTUS-TAG.
In addition, they both incorporate CoT reasoning, and Palimpzest also uses a more effective prompt with detailed instructions.
These design choices contribute to Palimpzest’s strong performance (70\% accuracy with Qwen3-Max), approaching our best-practice results
but at substantially higher cost (2--5$\times$ compared to our best-practice approach).
In contrast, HQDL relies on a single general-purpose LRO \texttt{Generate} to handle all tasks, capping its accuracy at around 30\%. 
HQDL also incurs substantially higher cost (about 10$\times$ than our best-practice approach), 
suggesting that an overly general, one-size-fits-all LRO is neither effective nor cost-saving.
Last, despite limited query coverage (41.7\%) and LRO coverage (2 LROs), 
ThalamusDB implements them following our best practices (\ie, \texttt{LLM-ONE} for \texttt{NLFILTER} and \texttt{LLM-ALL} for \texttt{NLJOIN}) and reaches 30\% peak accuracy.
ThalamusDB's accuracy is comparable to HQDL's, despite HQDL’s higher query coverage (75\%), highlighting that strong LRO designs can compensate for limited coverage.

Our best-practice baseline, with both comprehensive LRO coverage and  experimentally validated LRO designs, 
achieves the best performance across all base models, further demonstrating the importance of LRO support and design.

\insight{10}{\textbf{LRO Support and Design Effects.} 
Systems with broader coverage of our LRO taxonomy tend to achieve better performance.
Moreover, LRO implementations that follow our best practices often perform better, with CoT and instruction-rich prompts providing further gains.
In contrast, overly general, one-size-fits-all LROs are neither effective nor cost-efficient. 
}

\subsubsection{Performance \vs Base Model}
Base model choice also matters, but no single model always outperforms.
Due to different LRO implementation strategies, prompt styles, and context designs, systems exhibit various preferences for base models. 
GPT-5 performs best for SUQL, ThalamusDB, BlendSQL, LOTUS-TAG, and our best-practice baseline, 
while Qwen3-Max is more suitable for Palimpzest, 
and Claude-Sonnet-4.5 works best for HQDL and DocETL.
Besides performance, different base model choices also impact the monetary costs. 
Claude-Sonnet-4.5 is more expensive, due to  higher per-token pricing and greater reasoning token usage, 
whereas Qwen3-Max is cheaper with lower per-token pricing.


\insight{11}{\textbf{Base Model Effects.} 
Different base models substantially affect system performance and monetary cost.
However, higher-priced models are not necessarily better, and no single model is universally the best.
}

\begin{table*}[t]
    \centering
    \small
    \caption{Statistics of failure reasons (w.r.t. GPT-5) for end-to-end evaluation.}
    \label{tab:e2e-failure}
    \resizebox{\textwidth}{!}{
    \begin{tabular}{c|c|ccc|ccccccc} 
    \toprule
    \cellcolor{mygrey} & \cellcolor{mygrey} & \multicolumn{3}{c|}{\cellcolor{mygrey}\textbf{LLM-Planning Systems}} & \multicolumn{7}{c}{\cellcolor{mygrey}\textbf{Manual-Planning Systems}}      \\
    \cellcolor{mygrey} & \cellcolor{mygrey} & \cellcolor{mygrey}  & \cellcolor{mygrey}  & \cellcolor{mygrey}  & \cellcolor{mygrey}  & \cellcolor{mygrey}  & \cellcolor{mygrey}  & \cellcolor{mygrey} & \cellcolor{mygrey} & \cellcolor{mygrey} & \cellcolor{mygrey}  \\ 
    \multirow{-3}{*}{\cellcolor{mygrey}\textbf{Stage}} & \multirow{-3}{*}{\cellcolor{mygrey}\begin{tabular}[c]{@{}c@{}}\textbf{Failure}\\ \textbf{Reason}\end{tabular} } & \multirow{-2}{*}{\cellcolor{mygrey} Binder}  & \multirow{-2}{*}{\cellcolor{mygrey} SUQL} & \multirow{-2}{*}{\cellcolor{mygrey} \begin{tabular}[c]{@{}c@{}}Aryn-\\[-0.1em] Sycamore\end{tabular}}  & \multirow{-2}{*}{\cellcolor{mygrey} HQDL} & \multirow{-2}{*}{\cellcolor{mygrey} ThalamusDB}   &\multirow{-2}{*}{\cellcolor{mygrey} BlendSQL} & \multirow{-2}{*}{\cellcolor{mygrey} Palimpzest} & \multirow{-2}{*}{\cellcolor{mygrey} DocETL}  & \multirow{-2}{*}{\cellcolor{mygrey} LOTUS-TAG} & \multirow{-2}{*}{\cellcolor{mygrey} \begin{tabular}[c]{@{}c@{}}\textbf{Our Best}\\[-0.1em] \textbf{Practices} \end{tabular}} \\ 
    \midrule
    \textbf{Scope} & Out-of-scope & 68.97\% & 28.85\% & 66.67\%  & 38.10\%  & 87.50\% & -- & -- & --  & -- & -- \\
    \midrule
    \multirow{2}{*}{\textbf{Plan}} & Syntax error & 6.89\% & -- & -- & -- & -- & -- & -- & -- & -- & --\\
    & Plan logic error & 20.69\% & 67.30\% & 21.67\% & -- & -- & -- & -- & -- & -- & --\\
    \midrule
    \multirow{4}{*}{\textbf{Exec}} 
    & LRO misalignment\footnotemark[1] & --  & -- & -- & 14.29\% & -- & 37.04\% & 20.59\% & 3.85\% & 23.33\% & --\\
    & LLM hallucination & --  & 3.85\% & -- & -- & 7.50\% &  62.96\% & 79.41\% & 84.61\% & 70.00\%  & 100.00\% \\
    & Execution timeout  & 3.45\%  & -- & 11.66\% & 47.62\%  & 5.00\% & -- & -- & 11.54\% & 6.67\% & -- \\
    \bottomrule
    \end{tabular}
    }
\end{table*}

\subsubsection{Failure Analysis}
\label{sec:e2e-res-fa}

We further analyze the detailed failure reasons for each baseline. Table~\ref{tab:e2e-failure} reports six distinct failure modes identified in our analysis, spanning the scope, planning, and execution stages, where each numerical value represents the proportion of failures attributed to that mode relative to the total number of failures.

For LLM-planning systems, failures are mostly attributed to scope limitations and planning errors.
Binder and Aryn-Sycamore suffer primarily from scope limitations, as they only support single-table queries, leaving 40 queries out of scope each.
SUQL is heavily bottlenecked by planning errors, \ie, 67.30\% of failures are attributed to plan logic errors, stemming from its ambiguous planning instructions and one-size-fits-all LRO design (\ie, \texttt{Answer}), causing confusion for the LLM planner when deciding how and when to invoke them. 
This further confirms \iref{10}.
\footnotetext[1]{The implicit implementation of certain LLM-enhanced operators hinders correct results.}

For manual-planning systems, failures are confined to the scope limitation and execution stage, as their plans are handcrafted and thus error-free. 
As in Table~\ref{tab:correspond}, HQDL and ThalamusDB cover 75\% and 42\% queries, respectively, 
and therefore have a non-trivial number of out-of-scope failures.

Regarding LRO misalignment, HQDL, BlendSQL, Palimpzest, and LOTUS-TAG exhibit a non-negligible number of query failures attributable to this issue. 
The first three lack native support for \textsf{Cluster} and \textsf{Order} LROs, instead implicitly relying on \texttt{Generate}, \texttt{LLMMap}, and \texttt{Map} to generate new columns for cluster names or rankings, as illustrated in Table~\ref{tab:correspond}. 
LOTUS-TAG relies on clustering embedding vectors for its \texttt{sem\_cluster\_by}, which also results in a high misalignment failure rate.

Regarding LLM hallucination, DocETL, LOTUS-TAG, and BlendSQL exhibit substantial query failure rates of 84.61\%, 70.00\%, and 62.96\%, respectively.
Even our best-practice baseline is exclusively susceptible to LLM hallucination, confirming that hallucination is an inevitable challenge. 
Such hallucination is largely attributed to prompt design and base model capabilities, and thus warrants further investigation.

Execution timeout severely affects HQDL (47.62\%), as its full-table augmentation causes excessive token consumption, as discussed in Section~\ref{sec:e2e-lro-design}.
LOTUS-TAG and ThalamusDB also incur moderate timeout rates of 11.54\% and 6.67\%, respectively.

\insight{12}{
\textbf{Major Failure Modes.} 
We identify three major failure modes: (i) fragile planning cascading into downstream failures; (ii) suboptimal execution from misaligned or time-consuming LRO designs; and (iii) inevitable hallucination persisting across all systems.
}

\section{Conclusion and Outlook}
\label{sec:conclusion}

In this paper, we establish the first unified taxonomy for LLM-enhanced relational operators, 
organized along operating intents, operand granularities and implementation variants.
We design a comprehensive benchmark \proposal, 
spanning diverse real-world domains, with both single and multi-LRO workloads. 
We evaluate individual LROs and multi-LRO systems, deriving practical insights.
For individual LROs, we derive concrete LRO design choices (\iref{1}--\iref{5}) and trade-offs (\iref{6}--\iref{7}).
For systems, we highlight their early-stage landscape (\iref{8}),  our best-practice guidelines (\iref{9}), 
key performance factors: LRO support, LRO design, base model (\iref{10}--\iref{11}), 
and major failure modes (\iref{12}).


These insights suggest the need for a more principled design of LROs and their supporting systems.
In the short term, 
we expect LRO systems with: 
comprehensive LRO sets, 
robust LLM planning via reinforcement learning or prompt optimization, 
joint optimization of classical SQL and enhanced operators, 
and optimal LRO implementations with quality--cost trade-offs.

In the long term, we envision \emph{LLM-centric databases}, where databases are rebuilt around LLM-native operators, rather than having LLMs serve as plugins. 
Such databases would integrate: 
(i) multi-modal unified storage for heterogeneous data management; 
(ii) semantic-aware indexing beyond traditional or vanilla RAG mechanisms; 
(iii) adaptive context management via dynamically constructed prompt contexts; 
and (iv) holistic optimization jointly orchestrating storage, indexing, planning, and model invocation. 
At the core of this paradigm are LROs, serving as the abstraction layer that embeds LLM capabilities into query processing. 
This work establishes the first unified LRO taxonomy, a comprehensive benchmark \proposal, and empirical best-practice LRO designs to facilitate future research and system development.
\appendix

\section*{Appendix: Full Evaluation Results}

Full single-LRO evaluation results of \textsf{Filter}, \textsf{Match}, \textsf{Impute}, \textsf{Cluster} and \textsf{Order} are reported from Table~\ref{app:select} to Table~\ref{app:order}.

\begin{table*}[h]
\centering
\caption{Full evaluation results on \texttt{Filter}. (Best F1 scores in \textbf{bold}, second-best in \underline{underline} for each operand.)}
\label{app:select}
\resizebox{\textwidth}{!}{
\begin{tabular}{c |c | cccc | cccc |cccc} 
\toprule
\rowcolor{mygrey}
&   & \multicolumn{4}{c|}{\textbf{GPT-5}} & \multicolumn{4}{c|}{\textbf{Qwen3-Max}} & \multicolumn{4}{c}{\textbf{Claude-Sonnet-4.5}} \\
\rowcolor{mygrey}
\multirow{-2}{*}{\textbf{Operand}} & \multirow{-2}{*}{\textbf{Impl}} & \textbf{Precision} & \textbf{Recall} & \textbf{F1} & \textbf{Cost (\$)}  & \textbf{Precision} & \textbf{Recall} & \textbf{F1} & \textbf{Cost (\$)} & \textbf{Precision} & \textbf{Recall} & \textbf{F1} & \textbf{Cost (\$)} \\
 \midrule
\multirow{4}{*}{\begin{tabular}[c]{@{}c@{}}Row-wise\\ \texttt{Filter}\end{tabular}} 
 & \texttt{LLM-ALL} & 0.946 & 0.946 & 0.938 & 0.004 & 0.812 & 0.918 & 0.826 & 0.003 & 0.801 & 0.877 & 0.807 & 0.006\\
 &  {\footnotesize +CoT} & 0.947 & 0.957 & \textbf{0.945} & 0.005 & 0.946 & 0.909 & 0.917 & 0.004 & 0.851 & 0.881 & 0.846 & 0.009\\
 & \texttt{LLM-ONE} & 0.934 & 0.951 & 0.938 & 0.008 & 0.744 & 0.934 & 0.798 & 0.008 & 0.682 & 0.789 & 0.722 & 0.017\\
 & {\footnotesize +CoT} & 0.936 & 0.953 & \underline{0.940} & 0.025 & 0.917 & 0.962 & 0.932 & 0.026 & 0.931 & 0.961 & 0.937 & 0.078\\
 \midrule
 \multirow{4}{*}{\begin{tabular}[c]{@{}c@{}}Column-wise\\ \texttt{Filter}\end{tabular}} 
 & \texttt{LLM-ALL} & 0.947 & 0.841 & 0.851 & 0.001 & 0.941 & 0.906 & \textbf{0.903} & 0.001 & 0.941 & 0.802 & 0.826 & 0.003\\
 & {\footnotesize +CoT} & 0.938 & 0.895 & \underline{0.891} & 0.002 & 0.966 & 0.840 & 0.874 & 0.002 & 0.931 & 0.856 & 0.865 & 0.005\\
 & \texttt{LLM-ONE} & 0.845 & 0.921 & 0.872 & 0.005 & 0.866 & 0.896 & 0.854 & 0.005 & 0.782 & 0.932 & 0.830 & 0.011\\
 & {\footnotesize +CoT} & 0.860 & 0.893 & 0.849 & 0.014 & 0.803 & 0.994 & 0.869 & 0.015 & 0.781 & 0.820 & 0.771 & 0.043\\
  \midrule
 \multirow{4}{*}{\begin{tabular}[c]{@{}c@{}}Table-wise\\ \texttt{Filter}\end{tabular}} 
 & \texttt{LLM-ALL} & 0.914 & 0.928 & 0.904 & 0.009 & 0.929 & 0.955 & 0.928 & 0.009 & 0.937 & 0.943 & 0.930 & 0.021\\
 & {\footnotesize +CoT} & 0.946 & 0.919 & 0.921 & 0.009 & 0.945 & 0.957 & \underline{0.939} & 0.010 & 0.921 & 0.983 & \textbf{0.943} & 0.024\\
 & \texttt{LLM-ONE} & 0.725 & 0.678 & 0.670 & 0.011 & 0.674 & 0.604 & 0.606 & 0.011 & 0.646 & 0.716 & 0.664 & 0.025\\
 & {\footnotesize +CoT} & 0.689 & 0.678 & 0.654 & 0.020 & 0.675 & 0.613 & 0.612 & 0.021 & 0.612 & 0.664 & 0.617 & 0.060\\
\bottomrule
\end{tabular}
}
\end{table*}

\begin{table*}[h]
\centering
\caption{Full evaluation results on \textsf{Match}.
(Best F1 scores in \textbf{bold}, second-best in \underline{underline} for each operand.)}
\label{app:match}
\resizebox{\textwidth}{!}{
\begin{tabular}{c |c |cccc |cccc |cccc} 
\toprule
\rowcolor{mygrey}
 &    & \multicolumn{4}{c|}{\textbf{GPT-5}} & \multicolumn{4}{c|}{\textbf{Qwen3-Max}} & \multicolumn{4}{c}{\textbf{Claude-Sonnet-4.5}} \\
\rowcolor{mygrey}
\multirow{-2}{*}{\textbf{LRO}} & \multirow{-2}{*}{\textbf{Impl}} & \textbf{Precision} & \textbf{Recall} & \textbf{F1} & \textbf{Cost (\$)} & \textbf{Precision} & \textbf{Recall} & \textbf{F1} & \textbf{Cost (\$)} & \textbf{Precision} & \textbf{Recall} & \textbf{F1} & \textbf{Cost (\$)} \\

 \midrule
 \multirow{6}{*}{\begin{tabular}[c]{@{}c@{}}Cell-wise\\ \textsf{Match}\end{tabular}} & \texttt{LLM-ALL}         & 0.854 & 0.891 & \underline{0.838} & 0.004 & 0.767 & 0.742 & 0.715 & 0.003 & 0.757 & 0.807 & 0.757 & 0.006 \\
 & {\footnotesize +CoT}     & 0.858 & 0.880 & \textbf{0.843} & 0.004 & 0.839 & 0.783 & 0.758 & 0.003 & 0.836 & 0.871 & 0.821 & 0.008 \\
 & \texttt{LLM-ONE}         & 0.843 & 0.851 & 0.818 & 0.045 & 0.623 & 0.766 & 0.627 & 0.042 & 0.527 & 0.727 & 0.564 & 0.092 \\
 & {\footnotesize +CoT}     & 0.833 & 0.837 & 0.802 & 0.174 & 0.709 & 0.802 & 0.704 & 0.185 & 0.761 & 0.820 & 0.753 & 0.513 \\
 & \texttt{LLM-SEMI}        & 0.829 & 0.766 & 0.752 & 0.006 & 0.556 & 0.715 & 0.565 & 0.006 & 0.585 & 0.740 & 0.603 & 0.020 \\
 & {\footnotesize +CoT}     & 0.834 & 0.823 & 0.789 & 0.014 & 0.736 & 0.789 & 0.729 & 0.014 & 0.777 & 0.816 & 0.756 & 0.047 \\
\midrule
 \multirow{6}{*}{\begin{tabular}[c]{@{}c@{}}Row-wise\\ \textsf{Match}\end{tabular}} & \texttt{LLM-ALL}         & 0.975 & 0.956 & \textbf{0.961} & 0.027 & 0.804 & 0.792 & 0.795 & 0.019 & 0.949 & 0.951 & \underline{0.948} & 0.052 \\
 & {\footnotesize +CoT}     & 0.975 & 0.956 & \textbf{0.961} & 0.029 & 0.982 & 0.677 & 0.735 & 0.024 & 0.922 & 0.948 & 0.929 & 0.058 \\
 & \texttt{LLM-ONE}         & 0.846 & 0.822 & 0.829 & 0.211 & 0.874 & 0.839 & 0.844 & 0.190 & 0.959 & 0.839 & 0.869 & 0.505 \\
 & {\footnotesize +CoT}     & 0.846 & 0.822 & 0.829 & 0.338 & 0.886 & 0.854 & 0.862 & 0.335 & 0.846 & 0.774 & 0.797 & 0.946 \\
 & \texttt{LLM-SEMI}        & 0.949 & 0.929 & 0.934 & 0.074 & 0.658 & 0.751 & 0.690 & 0.070 & 0.906 & 0.929 & 0.914 & 0.161 \\
 & {\footnotesize +CoT}     & 0.949 & 0.929 & 0.934 & 0.085 & 0.916 & 0.945 & 0.922 & 0.079 & 0.944 & 0.951 & 0.941 & 0.200 \\
\midrule
  \multirow{6}{*}{\begin{tabular}[c]{@{}c@{}}Column-wise\\ \textsf{Match}\end{tabular}}  
 &   \texttt{LLM-ALL}      & 0.670 & 0.855 & \underline{0.737} & 0.003 & 0.662 & 0.746 & 0.693 & 0.002 & 0.373 & 0.351 & 0.360 & 0.006 \\
 & {\footnotesize +CoT}  & 0.685 & 0.846 & \textbf{0.746} & 0.003 & 0.736 & 0.771 & 0.735 & 0.003 & 0.702 & 0.772 & 0.726 & 0.008 \\
 & \texttt{LLM-ONE}      & 0.564 & 0.712 & 0.615 & 0.054 & 0.518 & 0.512 & 0.479 & 0.053 & 0.591 & 0.779 & 0.652 & 0.127 \\
 & {\footnotesize +CoT}  & 0.537 & 0.700 & 0.591 & 0.154 & 0.393 & 0.526 & 0.415 & 0.166 & 0.491 & 0.723 & 0.571 & 0.484 \\
 & \texttt{LLM-SEMI}       & 0.617 & 0.858 & 0.709 & 0.012 & 0.403 & 0.651 & 0.485 & 0.012 & 0.477 & 0.798 & 0.574 & 0.031 \\
 & {\footnotesize +CoT}  & 0.619 & 0.852 & 0.705 & 0.017 & 0.533 & 0.779 & 0.623 & 0.019 & 0.546 & 0.802 & 0.637 & 0.050 \\

\bottomrule
\end{tabular}
}
\end{table*}

\begin{table*}[h]
\centering
\caption{Full evaluation results on \textsf{Impute}.
(Best LLM judge scores in \textbf{bold}, second-best in \underline{underline} for each operand.)}
\label{app:impute}
\resizebox{\textwidth}{!}{
\begin{tabular}{c |c |ccc |ccc |ccc} 
\toprule
\rowcolor{mygrey}
 &  & \multicolumn{3}{c|}{\textbf{GPT-5}} & \multicolumn{3}{c|}{\textbf{Qwen3-Max}} & \multicolumn{3}{c}{\textbf{Claude-Sonnet-4.5}} \\
\rowcolor{mygrey}
\multirow{-2}{*}{\textbf{Operand}} & \multirow{-2}{*}{\textbf{Impl}}  & \textbf{EM} & \textbf{LLM Judge Score} & \textbf{Cost (\$)} & \textbf{EM} & \textbf{LLM Judge Score} & \textbf{Cost (\$)} & \textbf{EM} & \textbf{LLM Judge Score} & \textbf{Cost (\$)} \\
 \midrule
\multirow{4}{*}{\begin{tabular}[c]{@{}c@{}}Cell-wise\\ \textsf{Impute}\end{tabular}} 
 & \texttt{LLM-ALL}            & 0.733 & \textbf{0.890} & 0.003  & 0.643 & 0.826 & 0.003  & 0.661 & 0.864 & 0.007 \\
 & {\footnotesize +CoT}       & 0.739 & \underline{0.886} & 0.003  & 0.648 & 0.820 & 0.003  & 0.665 & 0.864 & 0.008 \\
 & \texttt{LLM-ONE}        & 0.691 & 0.836 & 0.001  & 0.576 & 0.746 & 0.001  & 0.635 & 0.792 & 0.004 \\
 & {\footnotesize +CoT}     & 0.674 & 0.848 & 0.002  & 0.596 & 0.824 & 0.003  & 0.634 & 0.833 & 0.008 \\
\midrule

\multirow{4}{*}{\begin{tabular}[c]{@{}c@{}}Column-wise\\ \textsf{Impute}\end{tabular}} 
 & \texttt{LLM-ALL}            & 0.606 & 0.807 & 0.001  & 0.557 & 0.740 & 0.001  & 0.606 & 0.817 & 0.002 \\
 & {\footnotesize +CoT}       & 0.577 & 0.816 & 0.001  & 0.526 & 0.737 & 0.001  & 0.610 & 0.834 & 0.003 \\
 & \texttt{LLM-ONE}            & 0.618 & 0.826 & 0.002  & 0.659 & 0.805 & 0.002  & 0.644 & 0.812 & 0.005 \\
 & {\footnotesize +CoT}       & 0.640 & \underline{0.838} & 0.005  & 0.621 & 0.793 & 0.006  & 0.657 & \textbf{0.859} & 0.017 \\
\midrule

\multirow{4}{*}{\begin{tabular}[c]{@{}c@{}}Row-wise\\ \textsf{Impute}\end{tabular}} 
 & \texttt{LLM-ALL}        & 0.674 & \underline{0.754} & $7.0\times 10^{-5}$ & 0.667 & \textbf{0.764} & $7.0\times 10^{-5}$ & 0.664 & 0.719 & $1.9\times 10^{-4}$ \\
 & {\footnotesize +CoT}    & 0.674 & 0.724 & $1.5\times 10^{-4}$ & 0.662 & 0.751 & $2.1\times 10^{-4}$ & 0.659 & 0.714 & $5.2\times 10^{-4}$ \\
 & \texttt{LLM-ONE}            & 0.438 & 0.632 & $7.0\times 10^{-5}$ & 0.447 & 0.644 & $8.0\times 10^{-5}$ & 0.455 & 0.580 & $1.9\times 10^{-4}$ \\
 & {\footnotesize +CoT}       & 0.422 & 0.607 & $1.5\times 10^{-4}$ & 0.479 & 0.664 & $1.9\times 10^{-4}$ & 0.433 & 0.561 & $4.8\times 10^{-4}$ \\
\bottomrule
\end{tabular}}
\end{table*}%

\thispagestyle{plain}

\begin{table*}[h]
\centering
\caption{Full evaluation results on \textsf{Cluster}.
(Best ARI, NMI in \textbf{bold}, second-best in \underline{underline} for each operand.)}
\label{app:cluster}
\resizebox{0.7\textwidth}{!}{
\begin{tabular}{c |c|ccc |ccc |ccc} 
\toprule
\rowcolor{mygrey}
 &  & \multicolumn{3}{c|}{\textbf{GPT-5}} & \multicolumn{3}{c|}{\textbf{Qwen3-Max}} & \multicolumn{3}{c}{\textbf{Claude-Sonnet-4.5}} \\
\rowcolor{mygrey}
\multirow{-2}{*}{\textbf{Operand}} & \multirow{-2}{*}{\textbf{Impl}}  & \textbf{ARI} & \textbf{NMI} & \textbf{Cost (\$)} & \textbf{ARI} & \textbf{NMI} & \textbf{Cost (\$)} & \textbf{ARI} & \textbf{NMI} & \textbf{Cost (\$)} \\
 \midrule
\multirow{4}{*}{\begin{tabular}[c]{@{}c@{}}Row-wise\\ \textsf{Cluster}\end{tabular}} & \texttt{LLM-ALL}  & \textbf{0.736} & \textbf{0.809} & 0.007 & \underline{0.731} & \underline{0.806} & 0.005 & 0.674 & 0.768 & 0.013 \\
 & {\footnotesize +CoT} & 0.723 & 0.801 & 0.008 & 0.720 & 0.778 & 0.006 & 0.675 & 0.763 & 0.016 \\
 & \texttt{LLM-ONE}  & 0.694 & 0.754 & 0.017 & 0.684 & 0.746 & 0.015 & 0.688 & 0.757 & 0.038 \\
 & {\footnotesize +CoT} & 0.694 & 0.759 & 0.029 & 0.658 & 0.727 & 0.026 & 0.696 & 0.756 & 0.073 \\
\midrule

\multirow{4}{*}{\begin{tabular}[c]{@{}c@{}}Column-wise\\ \textsf{Cluster}\end{tabular}} 
 & \texttt{LLM-ALL} & 0.822 & 0.890 & 0.009 & 0.807 & 0.869 & 0.007 & \underline{0.843} & \textbf{0.900} & 0.016 \\
 & {\footnotesize +CoT} & 0.812 & 0.877 & 0.010 & 0.751 & 0.835 & 0.008 & \textbf{0.848} & \underline{0.897} & 0.018 \\
 & \texttt{LLM-ONE} & 0.598 & 0.720 & 0.017 & 0.549 & 0.694 & 0.015 & 0.640 & 0.735 & 0.037 \\
 & {\footnotesize +CoT} & 0.604 & 0.721 & 0.028 & 0.493 & 0.641 & 0.027 & 0.610 & 0.726 & 0.077 \\
\midrule

\multirow{4}{*}{\begin{tabular}[c]{@{}c@{}}Table-wise\\ \textsf{Cluster}\end{tabular}} 
 & \texttt{LLM-ALL} & 0.699 & 0.836 & 0.015 & \textbf{0.743} & \textbf{0.852} & 0.014 & 0.707 & 0.831 & 0.033 \\
 & {\footnotesize +CoT} & \underline{0.721} & \underline{0.837} & 0.015 & 0.627 & 0.783 & 0.015 & 0.716 & 0.830 & 0.035 \\
 & \texttt{LLM-ONE} & 0.679 & 0.810 & 0.017 & 0.626 & 0.782 & 0.016 & 0.623 & 0.771 & 0.040 \\
 & {\footnotesize +CoT} & 0.668 & 0.801 & 0.023 & 0.630 & 0.780 & 0.023 & 0.647 & 0.788 & 0.059 \\
\bottomrule
\end{tabular}
}
\end{table*}%

\thispagestyle{plain}

\begin{table*}[h]
\centering
\caption{Full evaluation results on \textsf{Order}. (Best HR@$k$, Kendall $\tau$ in \textbf{bold}, second-best in \underline{underline}.)}
\label{app:order}
\resizebox{0.8\textwidth}{!}{
\begin{tabular}{c |c |ccc |ccc |ccc} 
\toprule
\rowcolor{mygrey}
 &  & \multicolumn{3}{c|}{\textbf{GPT-5}} & \multicolumn{3}{c|}{\textbf{Qwen3-Max}} & \multicolumn{3}{c}{\textbf{Claude-Sonnet-4.5}} \\
\rowcolor{mygrey}
\multirow{-2}{*}{\textbf{Operand}} & \multirow{-2}{*}{\textbf{Impl}}    & \textbf{HR@$k$} & \textbf{Kendall $\tau$} & \textbf{Cost (\$)} & \textbf{HR@$k$} & \textbf{Kendall $\tau$} & \textbf{Cost (\$)} & \textbf{HR@$k$} & \textbf{Kendall $\tau$} & \textbf{Cost (\$)}  \\
\midrule
\multirow{8}{*}{\begin{tabular}[c]{@{}c@{}}Row-wise\\ \textsf{Order}\end{tabular}}
& \texttt{LLM-ALL}        & 0.929 & 0.899 & 0.001 & 0.861 & 0.693 & 0.001 & \underline{0.945} & 0.848 & 0.003 \\
& {\footnotesize +CoT}    & 0.941 & \textbf{0.937} & 0.002 & 0.952 & 0.884 & 0.002 & \textbf{0.949} & 0.865 & 0.006 \\
& \texttt{LLM-PAIR}       & 0.929 & 0.905 & 0.067 & 0.901 & 0.757 & 0.067 & 0.767 & 0.688 & 0.133 \\
& {\footnotesize +CoT}    & 0.929 & 0.907 & 0.182 & 0.921 & 0.892 & 0.208 & 0.910 & \underline{0.914} & 0.568 \\
& \texttt{LLM-SORT}       & 0.885 & 0.867 & 0.008 & 0.901 & 0.858 & 0.011 & 0.828 & 0.816 & 0.024 \\
& {\footnotesize +CoT}    & 0.896 & 0.881 & 0.021 & 0.941 & 0.884 & 0.033 & 0.899 & 0.871 & 0.083 \\
& \texttt{LLM-SCORE}      & 0.789 & 0.690 & 0.004 & 0.831 & 0.686 & 0.004 & 0.807 & 0.707 & 0.009 \\
& {\footnotesize +CoT}    & 0.776 & 0.658 & 0.011 & 0.805 & 0.588 & 0.014 & 0.798 & 0.706 & 0.039 \\
\bottomrule
\end{tabular}
}
\end{table*}%

\clearpage


\clearpage

\bibliographystyle{spmpsci}
\bibliography{ref}

\clearpage

\end{document}